\documentclass[runningheads]{llncs}
\usepackage[T1]{fontenc}
\usepackage{graphicx}
\usepackage[misc]{ifsym}
\newcommand{\corr}{(\Letter)}
\graphicspath{{./}{figs/}}
\DeclareGraphicsExtensions{.pdf,.png,.jpeg,.jpg}

\usepackage{amsmath,amssymb}
\usepackage{booktabs}
\usepackage{multirow}
\usepackage{makecell}
\usepackage{rotating}
\usepackage{url}
\usepackage[hidelinks]{hyperref}
\usepackage[capitalize,nameinlink]{cleveref}
\usepackage{xspace}
\usepackage{xcolor}
\usepackage{subcaption}
\usepackage{diagbox}
\usepackage{adjustbox}
\usepackage{multibib}
\newcites{app}{Online Appendix References}
\usepackage{pgfplots}
\pgfplotsset{compat=1.18}
\usetikzlibrary{patterns}

\definecolor{navy}{rgb}{0.1, 0.1, 0.8}
\definecolor{gray}{rgb}{0.4, 0.4, 0.4}
\definecolor{olive}{rgb}{0.1, 0.5, 0.1}
\definecolor{ruby}{rgb}{0.8, 0.1, 0.3}
\definecolor{darkpastelgreen}{rgb}{0.01, 0.75, 0.24}
\definecolor{celestialblue}{rgb}{0.29, 0.59, 0.82}
\definecolor{coral}{rgb}{1.0, 0.5, 0.31}
\definecolor{blue}{rgb}{0.23, 0.44, 0.62}
\definecolor{Goldenrod}{rgb}{0.8,0.8,0}

\usepackage[colorinlistoftodos,textsize=tiny]{todonotes} \usepackage{soul}

\newcommand{\eat}[1]{}

\newcommand{\NOTE}[2]{}
\newcommand{\editnote}[2][1=]{}
\newcommand{\nb}[1]{}
\newcommand{\mar}[1]{}
\newcommand{\roh}[1]{}
\newcommand{\TODO}[2]{}

\newcommand{\rbrexit}{\textsc{r/Brexit}\xspace}

\begin{document}

\title{Brexit Means Brexit: Selection Bias, Echo Chambers, and Entrenched Opinion on Reddit}

\titlerunning{Polarisation Dynamics on Structured Discussion Platforms}

\author{
  Marian-Andrei Rizoiu\inst{1} \corr \and
  Duy Khuu\inst{1} \and
  Andrew Law\inst{1} \and
  Christine Largeron\inst{2}
}

\authorrunning{M.-A.\ Rizoiu et al.}

\institute{
  University of Technology Sydney, Australia\\
  \url{www.behavioral-ds.science}\\
  \email{Marian-Andrei.Rizoiu@uts.edu.au} \and
  Universit\'{e} de Lyon, UJM-Saint-\'{E}tienne, CNRS, France
}

\toctitle{Brexit Means Brexit: Selection Bias, Echo Chambers, and Entrenched Opinion on Reddit}
\tocauthor{Marian-Andrei Rizoiu, Duy Khuu, Andrew Law, Christine Largeron}

\maketitle

\begin{abstract}
Political polarisation on structured discussion platforms such as Reddit differs fundamentally from that on broadcast platforms such as Twitter/X, yet most prior work targets the latter.
We present an end-to-end framework for measuring and analysing polarisation dynamics, applied to the \rbrexit subreddit (871K submissions, November 2015 -- February 2021).
We construct \textbf{\rbrexit}, a crowd-annotated stance dataset of 5{,}024 labelled submissions (inter-annotator agreement~$= 0.804$), and train a domain-adapted BERT classifier.
We introduce a continuous polarity metric that replaces discrete stance categories, revealing fine-grained opinion spectra across 27 politically-defined periods.
Our analysis yields three findings:
(a)~future stance prediction is confounded by survivorship bias: who remains active is self-selected on engagement, not stance, biasing any longitudinal model toward a non-representative minority;
(b)~echo chambers are quantifiably dominant, with nearly 40\% of interactions between like-minded users;
(c)~user current polarity is the dominant predictor of future polarity, with echo-chamber immersion as the secondary predictive signal.
These findings reveal that Reddit's partisan core is entrenched by self-selection, not softened by cross-cutting exposure.

\keywords{Polarisation dynamics \and Stance detection \and Echo chambers \and Reddit \and Continuous polarity \and Selection bias}
\end{abstract}

\section{Introduction}
\label{sec:intro}

\textbf{Does engagement with opposing views change political opinions online?} 
On broadcast platforms such as Twitter/X, the question is largely moot: opposing camps cluster around distinct hashtags, retweets signal endorsement, and users rarely encounter dissent~\cite{aldayel2021stance,conover2011political}.
Structured discussion forums (Reddit chief among them) offer a starkly different environment: opposing users reply within the same thread, pseudonymity lowers social cost, and long-form deliberation is the structural norm~\cite{bruns2017echo}.
If cross-cutting exposure can shift opinions anywhere, it should happen on Reddit.
We find that on \rbrexit, it largely does not, and the reason reveals a structural confound prior work has overlooked.

Three questions must be answered to study this.
\textbf{Q1: Can reliable stance labels be obtained for Reddit political discussions?}
Prior work~\cite{Largeron2021} relied on a Twitter-trained stance classifier that agrees with crowd annotations on only ${\sim}25\%$ of comments, producing near-random performance (F1~$= 0.32$) on the Reddit domain.
\textbf{Q2: How should user-level political stance be represented from discrete post labels?}
Naive majority-vote aggregation collapses nuance: it labels 83\% of users as ``Neutral,'' making a user who posts two pro-Brexit and three neutral comments indistinguishable from one who posts exclusively neutral content, masking genuine opinion diversity.
\textbf{Q3: Does user attrition introduce a selection bias that confounds longitudinal stance analyses?}
Our observations are already restricted to \emph{active contributors}: Reddit's API exposes content but not reads, so lurkers are invisible.
Among active users, over 70\% of \rbrexit users appear in only one of 27 time periods, and fewer than 1\% sustain engagement across three consecutive periods, rendering ``future stance prediction'' conditional on the user returning and actively contributing again, a confounder unacknowledged in prior work~\cite{Kong2022,Largeron2021}.

We present an end-to-end framework that addresses all three questions, applied to the \rbrexit subreddit spanning the full Brexit saga (November 2015 -- February 2021; 871K submissions).
The answer to \textbf{Q3} reframes the other two: the dominant mechanism shaping polarisation dynamics on \rbrexit is not opinion change but \emph{survivorship bias}.
The active community is overwhelmingly transient, and who keeps posting is decided by engagement, not conviction: the most strident voices are disproportionately one-off posters who never return.
This finding challenges the premise of prior future-stance prediction models and carries direct implications for platform polarisation interventions.

Our contributions are: 

\begin{enumerate}
\item \textbf{\rbrexit dataset.}
  A crowd-annotated Reddit stance dataset of 5{,}024 labelled submissions spanning 27 time periods, with inter-annotator agreement (IAA~$= 0.804$).
  We detail a systematic MTurk methodology including malicious worker detection.\footnote{The \rbrexit dataset and analysis code are available at \url{https://github.com/behavioral-ds/brexit-opinion-dynamics}.}

\item \textbf{Continuous polarisation measurement.}
  A domain-adapted BERT stance classifier (BERT-Reddit; macro F1~$= 0.555$) provides per-comment stance labels; these are aggregated via a continuous polarity metric that captures the full opinion spectrum beyond the three discrete categories.

\item \textbf{Behavioural findings.}
  (a)~Future stance prediction is fundamentally confounded by selection bias in user retention: who returns is governed by engagement, not stance: models condition on a non-representative, highly-engaged minority.
  (b)~Echo chambers are quantifiably dominant: nearly 40\% of user--period interactions occur between like-minded individuals.
  (c)~Among persistent users, opinion is stable: SHAP analysis identifies current polarity as the dominant predictor of future polarity, with echo-chamber immersion as the secondary signal.
  (d)~Richer models cannot overcome this ceiling: augmenting activity features with stance triads, a graph attention network, or sequential models yields negligible gains, because self-selection sets a predictive boundary for network topology.
\end{enumerate}

\section{Related Work and Prerequisites}
\label{sec:related}

\paragraph{Stance detection.}
Stance detection has progressed from SVM classifiers with n-gram features~\cite{mohammad2016semeval} through network-augmented approaches~\cite{conover2011political} to transformer-based models, with BERT fine-tuning, augmented by domain-specific continued pre-training, now the standard approach~\cite{aldayel2021stance,alturayeif2023systematic,nguyen2020bertweet,sun2019fine}.
Cross-context transferability is a persistent challenge: signal choice critically determines pipeline portability across platforms and time periods~\cite{Ram2025}, and training across heterogeneous labelled corpora substantially improves cross-dataset generalisation under limited supervision~\cite{Yuan2023,Yuan2024}.

Zero-shot LLM prompting (via chain-of-thought~\cite{gatto2023cot} and counterfactual strategies~\cite{weinzierl2024tree}) has been shown to rival fine-tuned models on established benchmarks.
However, converging evidence shows that fine-tuned small language models outperform zero-shot LLMs when in-domain labelled data is available~\cite{Lee2026}, while parameter-efficient fine-tuning additionally yields interpretable, expert-grounded detectors for politically charged social media text~\cite{Tian2025XTroll}.
Ziems et al.~\cite{ziems2024llms} evaluate 13~LLMs across 25~computational social science tasks and find they do not outperform fine-tuned classifiers.
Furthermore, Gera and Neal~\cite{gera2025deep} confirm in a comprehensive survey that BERT variants remain dominant across stance detection settings.
We therefore adopt a fine-tuned BERT approach with domain-specific continued pre-training on our Reddit corpus (\cref{sec:stance_class}).

A gap remains: most stance detection research targets Twitter/X, whose short-form posts differ markedly from Reddit's tree-structured deliberative discussions~\cite{alturayeif2023systematic,gera2025deep}.
The few Reddit-specific efforts address multi-target user-level stance discovery~\cite{steel2024multi} or inter-user disagreement prediction~\cite{lorge2024stentconv}, but neither provides crowd-annotated stance labels for a single contentious topic at scale.
Our \rbrexit dataset fills this gap with 5{,}024 crowd-annotated texts, a systematic MTurk methodology with malicious-worker detection, and a domain-adapted BERT classifier evaluated against external human judgements.

\paragraph{Opinion dynamics.}
Computational models of opinion dynamics range from RNN-based sequential models of stance change~\cite{Kong2022} and sociologically-informed neural networks embedding classical models (DeGroot, bounded confidence) within deep learning frameworks~\cite{okawa2022sinn}, to variational inference approaches for agent-based model calibration~\cite{lenti2025variational}, with generative agent simulations reproducing emergent social ties and homophilous network structure from individual interaction rules~\cite{Schneider2025}.
Complementary work measures social influence via crowdsourced pairwise comparisons and psychosocial-inspired diffusion models~\cite{Ram2024,Ram2026}, disentangles diffusion pathways via Hawkes processes~\cite{Calderon2024b}, models engagement as a sequence of social-exchange decisions~\cite{Tian2026}, and detects emerging narratives via state-space models~\cite{Tian2025ICMamba}.
At the platform level, Cinus et al.~\cite{cinus2022recommenders} show that people recommenders amplify echo chambers only when substantial initial homophily already exists; otherwise algorithmic effects are negligible.
A common limitation is the assumption of known social neighbourhoods (e.g., Twitter follower graphs) and fixed user populations; on Reddit, networks must be inferred from reply structure, and the population is open, as participants arrive and depart freely.

On Reddit, opinion change alters subsequent community participation patterns~\cite{petruzzellis2023opinion}, yet prior studies overwhelmingly treat stance prediction \emph{unconditionally}: they predict a user's future stance without accounting for whether the user will remain present~\cite{Kong2022,Largeron2021}.
A field experiment by Oswald et al.~\cite{oswald2025disentangling} shows online political discussions are dominated by a systematically atypical active minority, so self-selection distorts observed opinion distributions.
No prior work formally quantifies how user attrition biases longitudinal polarisation estimates, nor jointly models presence and stance as interconnected outcomes.
We address both: we analyse which features predict departure versus opinion persistence (\cref{sec:presence}), and we identify survivorship bias as a fundamental confound in stance dynamics (\cref{sec:discussion}).

\begin{figure}[t]
  \centering
  \includegraphics[width=\columnwidth]{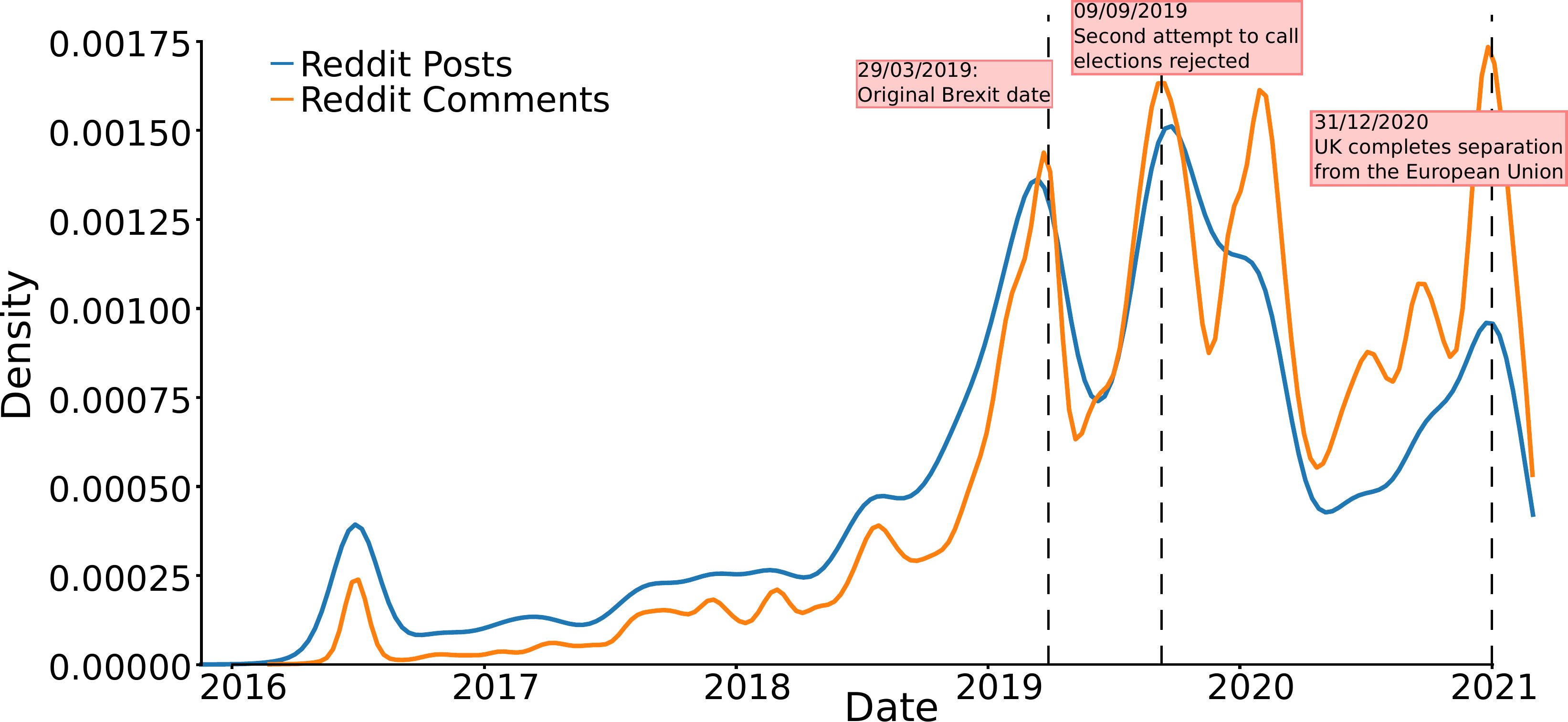}
  \caption{Temporal density of posts and comments in \rbrexit (November 2015 -- February 2021).
    Activity surges coincide with the original Brexit date (29/03/2019), the second attempt to call elections (09/09/2019), and the UK's final separation (31/12/2020).}
  \label{fig:temporal_density}
\end{figure}

\paragraph{Political polarisation on Reddit.}
Reddit has become a primary site for studying online political discourse, with dedicated resources such as the Reddit Politosphere covering 605 political subreddits~\cite{hofmann2022politosphere}.
Echo-chamber research reveals a more complex picture than simple polarisation narratives: hostile speech is more prevalent \emph{within} than between opposing sides~\cite{efstratiou2023nonpolar}, and polarised news sharing is concentrated in a small minority of users (``echo tunnels'' rather than broad chambers~\cite{mok2023echo}).
Segregation in news discussions is driven more by demographics than by ideology~\cite{monti2023demographic}, and a single left--right axis is insufficient: interactions on r/PoliticalCompass exhibit homophily on the social axis but heterophily on the economic axis~\cite{colacrai2024compass}.
Selective exposure operates even at the level of topic choice: fringe users disproportionately cherry-pick narrow subsets of otherwise mainstream content rather than consuming overtly false material~\cite{Lee2025facts}.
Users who deliberately engage with opposing views use less hostile language, yet their home communities penalise them with fewer upvotes~\cite{efstratiou2024deliberate}, which suggests structural disincentives for bridge-building.
At the ecosystem level, far-right opinions self-reinforce and compete with moderate views for finite attention, and targeted interventions can stem this spread~\cite{Calderon2024a}; content moderation itself measurably reduces harm when applied within tight time windows~\cite{Schneider2023}.
Yuan et al.~\cite{Yuan2025} extend homophily measurement via inverse reinforcement learning, showing users can share posting behaviour while discussing different topics, and identify a class of users oriented toward disagreement.

In our own prior work~\cite{Largeron2021}, we studied the same \rbrexit subreddit and introduced a feature-based framework for predicting a user's future stance given their activity in period~$t$.
We defined four feature sets, namely textual content (F0), user-level activity statistics (F1, F2), and discussion-composition descriptors (F3), which we refer to collectively as~F0123 (full feature inventory in Online Appendix~D~\cite{appendix}).
That work reported macro F1~$= 0.539$ for the best feature set (F3), but relied on a Twitter-trained Naive Bayes classifier for stance labels; 
here, we show (\cref{sec:profile}) that this classifier agrees with crowd annotations on only ${\sim}25\%$ of comments, predominantly because it over-predicts ``Neutral.''
Community-level analyses~\cite{Colleoni:2014} and earlier studies of online political discourse~\cite{wojcieszak2009online} document homophily at the platform level but do not measure \emph{within-subreddit} polarisation dynamics at the user level over time.
Our framework addresses this: we quantify echo chambers via edge homogeneity within a single subreddit, track individual users across 27~politically-segmented periods, and show that the users who persist are precisely those most resistant to opinion change.

\section{The \rbrexit Dataset}
\label{sec:dataset}

Studying polarisation dynamics at scale requires ground-truth stance labels, yet the 871K submissions in \rbrexit preclude exhaustive manual annotation.
We therefore construct a high-quality annotated subset: we temporally segment the corpus (\cref{sec:datacollect}), crowd-annotate a stratified sample via Amazon Mechanical Turk (\cref{sec:annotation}), producing the \rbrexit dataset profiled in \cref{sec:profile}.

\subsection{Data Collection and Platform Context}
\label{sec:datacollect}

Reddit is a pseudonymous platform organised into topic-specific communities (\emph{subreddits}), where posts initiate tree-structured comment threads, with branching multi-level deliberation distinct from flat broadcast platforms.

We originally introduce the \rbrexit dataset in our prior work~\cite{Largeron2021}; 
here, we reconstruct and extend it using the Pushshift API~\cite{pushshift}, spanning November 2015 through February 2021.\footnote{The Pushshift Reddit dataset service was discontinued in 2023; Reddit data collected prior to that date remains publicly downloadable.}
The corpus comprises 871{,}955 submissions (56{,}017 posts and 815{,}938 comments).
We segment the timeline into 27 periods demarcated by high-profile political events: from the referendum through Theresa May's resignation, Boris Johnson's ascent, and the UK's final separation from the EU; the full period table is in Online Appendix~A~\cite{appendix}.
Event-based segmentation ensures each period captures a coherent phase of the debate.

\Cref{fig:temporal_density} shows the temporal distribution of submissions.
Activity increases markedly from 2019 onward, with surges at major political events.

\subsection{Crowd-Sourced Stance Annotation}
\label{sec:annotation}

We annotate a subset of submissions using Amazon Mechanical Turk (MTurk)~\cite{mturkapi}.
Following Mohammad et al.~\cite{mohammad2016semeval}, we adopt three stance classes: \emph{pro-Brexit}, \emph{anti-Brexit}, and \emph{neither}.
Each submission is annotated by eight workers, with the majority label retained; workers see the full comment thread for context.
We achieve annotation quality through three steps: parameter optimisation on small test batches, malicious-worker detection in production, and a low-confidence filter on the final dataset.

\paragraph{Step 1 — Parameter optimisation.}
We run a series of test batches (200 submissions each), varying one parameter at a time.
\Cref{fig:mturk_params} summarises the effects.
Restricting workers to English-speaking countries improves IAA (inter-annotator agreement, i.e., the proportion of annotations matching the majority label) by 3.6\%; tightening prior-performance requirements (approval rate [AR]~$\geq 98\%$, minimum HITs completed [MHC]~$\geq 1{,}000$) adds a further 4.5\%.
The best configuration yields IAA~$= 0.798$, comparable to the $0.82$ reported by Mohammad et al.~\cite{mohammad2016semeval}.

\newsavebox{\mturkimg}\savebox{\mturkimg}{\includegraphics[width=0.58\columnwidth]{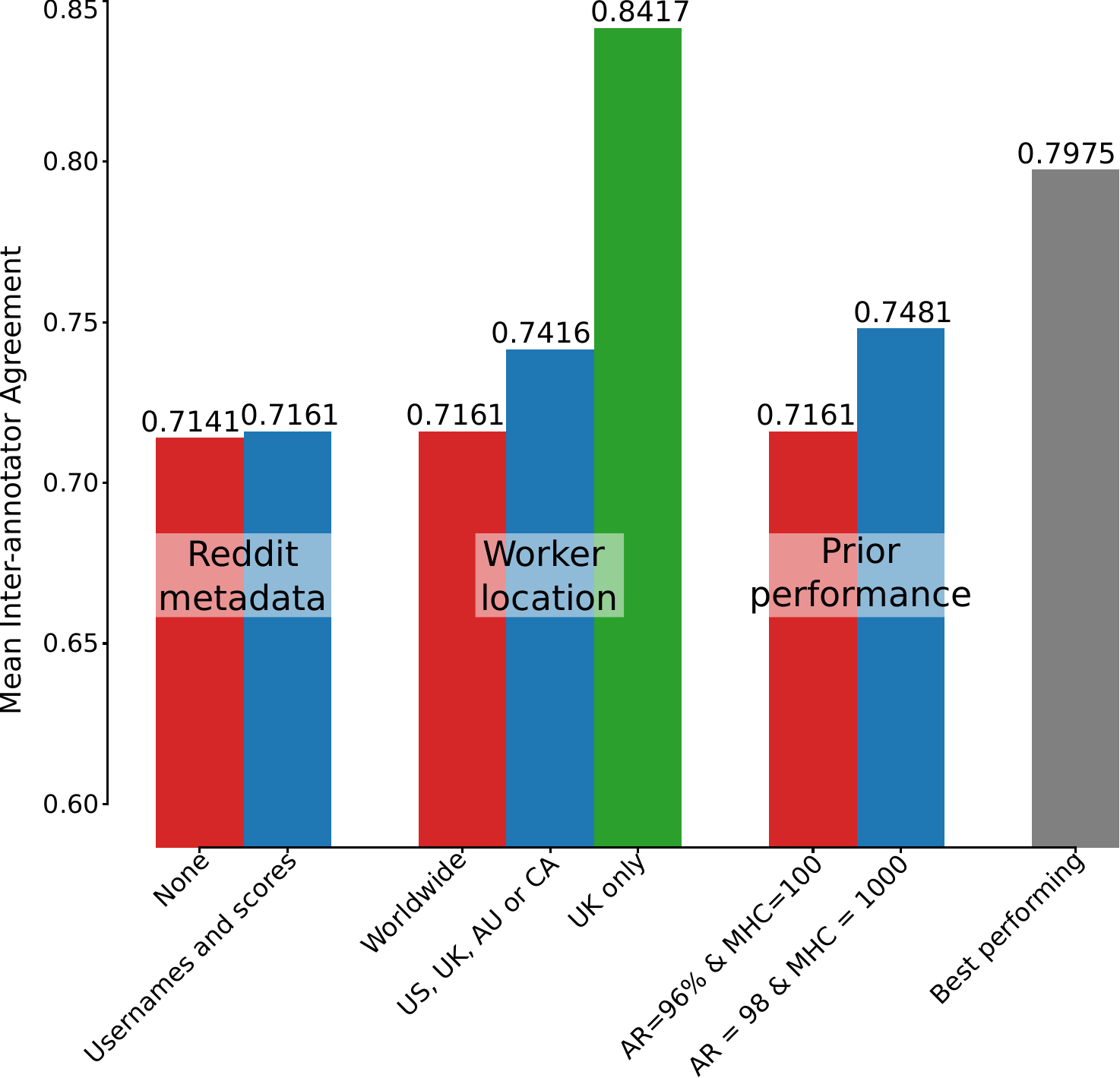}}\begin{figure}[tb]
  \centering
  \subfloat[\label{fig:mturk_params}]{\usebox{\mturkimg}}\hfill \subfloat[\label{tab:class_iaa}]{\begin{minipage}[c][\ht\mturkimg+\dp\mturkimg][c]{0.37\columnwidth}\centering
      \small\begin{tabular}{lcc}
        \toprule
        \textbf{Class} & \textbf{IAA (all)} & \textbf{IAA (filt.)} \\
        \midrule
        Pro-Brexit  & 0.623 & 0.720 \\
        Neither     & 0.759 & 0.824 \\
        Anti-Brexit & 0.694 & 0.742 \\
        \midrule
        \textbf{Overall} & \textbf{0.740} & \textbf{0.804} \\
        \bottomrule
      \end{tabular}\end{minipage}}\caption{
    \textbf{MTurk annotation quality.}
    \textbf{(a)}~Step~1: parameter sweep over worker qualifications (MHC: minimum HITs completed; AR: approval rate; full sweep in Online Appendix~B~\cite{appendix}); the best configuration yields IAA~$= 0.798$ (four English-speaking countries; UK-only is higher but causes unfinished batches).
    \textbf{(b)}~Step~3: per-class IAA before (``all'', post Step~2 allowlist, overall $0.740$) and after (``filt.'') the low-confidence filter (overall $0.804$).
  }
  \label{fig:mturk_iaa}
\end{figure}

\paragraph{Step 2 — Malicious worker detection.}
Despite optimal parameters, the first two production batches yielded IAA of $0.746$ and $0.710$, well below the expected $0.798$.
Inspection revealed a few prolific workers submitting annotations without reading the text, labelling randomly or with a fixed strategy to minimise effort.
Standard MTurk qualification requirements (approval rate~$\geq 98\%$, $\geq 1{,}000$ completed HITs) are insufficient to screen out such workers, since requesters are often reluctant to reject HITs.

To address this, we introduce the \emph{Majority Agreement Proportion} (MAP): the proportion of a worker's annotations that agree with the majority label of the remaining annotators.
Workers with $\text{MAP} < 0.25$ are flagged as malicious; re-annotating their submissions raises the two affected batch IAAs to $0.87$ and $0.91$, respectively.
An allowlist of workers with $\text{MAP} \geq 0.5$ and at least 20 prior annotations is then applied to all subsequent production batches, yielding an overall production IAA of $0.740$ (the ``IAA (all)'' column in~\cref{tab:class_iaa}).

\paragraph{Step 3 — Low-confidence filter.}
\label{sec:profile}
We discard submissions where fewer than five of eight annotators agree (IAA~$< 0.6$), as such cases signal genuine ambiguity rather than annotator error.
Removing these low-confidence instances raises the final dataset IAA to $0.804$ (the ``IAA (filt.)'' column in~\cref{tab:class_iaa}), the value used for all subsequent classifier training.
The final annotated dataset contains 5{,}024 labelled texts: 225 pro-Brexit (4.5\%), 887 anti-Brexit (17.7\%), and 3{,}912 neither (77.9\%).
Pro-Brexit exhibits the lowest per-class IAA ($0.72$), as workers encounter fewer exemplars of pro-Brexit language; neither is the easiest to annotate ($0.82$), containing less group-specific jargon.
Full details of the MTurk interface design, parameter tuning batches, and malicious-worker detection are provided in Online Appendix~B~\cite{appendix}.

\section{Measuring Continuous Polarisation}
\label{sec:method}

Our goal is to characterise each user's stance as a continuous quantity derived from the discrete labels of their individual posts.
This requires two steps: first, a stance classifier that assigns per-submission labels across the full corpus (\cref{sec:stance_class}); second, an aggregation scheme that maps these discrete labels to a continuous polarity score per user per period (\cref{sec:polarity}).

\subsection{Submission Stance Classification}
\label{sec:stance_class}

\paragraph{Models.}
Our prior work on \rbrexit~\cite{Largeron2021} relied on a Twitter-trained Naive Bayes classifier that agrees with our crowd annotations on only ${\sim}25\%$ of comments, predominantly because it over-predicts ``Neutral.''
We therefore train an in-domain classifier from scratch; fine-tuned SLMs consistently outperform zero-shot LLMs when in-domain labelled data is available~\cite{gera2025deep,Lee2026,ziems2024llms}, motivating a BERT fine-tuning approach.
We compare four strategies.
\emph{BERT-base} uses the standard pre-trained model without adaptation.
\emph{BERTweet}~\cite{nguyen2020bertweet} substitutes a model pre-trained on 850M English tweets.
\emph{BERT-DINO} uses DINO~\cite{schick2021generating} to generate over 4M synthetic \rbrexit-domain text pairs, then pre-trains BERT on semantic textual similarity before stance fine-tuning.
\emph{BERT-Reddit} (ours) continues masked-language-model (MLM) pre-training on all 871K \rbrexit submissions to adapt the model to subreddit vocabulary and discourse before fine-tuning.
All four are trained with randomised hyperparameter search and evaluated via stratified five-fold cross-validation (full hyperparameter spaces in Online Appendix~C~\cite{appendix}).

\begin{figure}[t]
  \centering
  \begin{tikzpicture}
    \begin{axis}[
      ybar,
      bar width      = 7pt,
      width          = 0.5\columnwidth,
      height         = 4cm,
      ymin=0, ymax=0.8,
      ytick          = {0, 0.2, 0.4, 0.6, 0.8},
      enlarge x limits = 0.46,
      symbolic x coords = {Accuracy, Macro-F1},
      xtick          = data,
      x tick label style = {font=\small},
      y tick label style = {font=\small, /pgf/number format/fixed,
                            /pgf/number format/precision=1},
      ylabel         = {Score},
      ylabel style   = {font=\small},
      axis line style = {draw=none},
      tick style      = {draw=none},
      legend style   = {at={(1.02,0.5)}, anchor=west,
                        legend columns=1, font=\scriptsize,
                        draw=none, fill=none},
      legend cell align = left,
      nodes near coords={\pgfmathprintnumber[fixed,precision=3]{\pgfplotspointmeta}},
      nodes near coords style = {font=\tiny, rotate=90, anchor=east,
                                 inner sep=2pt},
    ]
\addplot[fill=gray!40, draw=gray!60]
        coordinates {(Accuracy,0.545) (Macro-F1,0.320)};
\addplot[fill=blue!40, draw=blue!60]
        coordinates {(Accuracy,0.708) (Macro-F1,0.492)};
\addplot[fill=orange!50, draw=orange!70]
        coordinates {(Accuracy,0.779) (Macro-F1,0.487)};
\addplot[fill=cyan!40, draw=cyan!60]
        coordinates {(Accuracy,0.764) (Macro-F1,0.503)};
\addplot[fill=green!40, draw=green!60]
        coordinates {(Accuracy,0.745) (Macro-F1,0.550)};
\addplot[fill=violet!40, draw=violet!60]
        coordinates {(Accuracy,0.708) (Macro-F1,0.460)};
\addplot[fill=red!50, draw=red!70]
        coordinates {(Accuracy,0.783) (Macro-F1,0.555)};
      \legend{Twitter-stance~\cite{Largeron2021}, SVM (TF-IDF), XGBoost,
              BERTweet~\cite{nguyen2020bertweet},
              BERT-base~\cite{devlin2018bert},
              BERT-DINO~\cite{schick2021generating},
              BERT-Reddit (ours)}
    \end{axis}
  \end{tikzpicture}
  \caption{Stance classification results (stratified five-fold cross-validation).
           BERT-Reddit achieves the best macro-F1 ($0.555$) and accuracy ($0.783$).}
  \label{tab:results}
\end{figure}

\paragraph{Results.}
\Cref{tab:results} summarises classifier performance.
BERT-Reddit outperforms all alternatives, achieving macro-F1~$= 0.555$ against $0.320$ for the Twitter transfer baseline.
Notably, BERTweet underperforms BERT-base, suggesting Reddit language is closer to standard English than to Twitter vernacular.
BERT-DINO degrades performance, reflecting noise in the synthetically generated STS pairs.
F1~$= 0.555$ is contextually strong: the random-chance baseline for this heavily imbalanced dataset is $0.26$ (not $0.33$), as the pro-Brexit class constitutes only 5\% of submissions.
We use BERT-Reddit's predictions as stance labels throughout; the sensitivity of downstream findings to label error is bounded and contextualised in \cref{sec:discussion}.

\subsection{From Discrete Stance to Continuous Polarity}
\label{sec:polarity}

\paragraph{Continuous user polarisation score.}
Majority-vote aggregation of per-post stances collapses nuance: under it, 83\% of users are labelled ``Neutral,'' masking a substantially polarised population.
We instead define a \emph{continuous polarity} for user $u$ in period $t$ as:
\begin{equation}\label{eq:polarity}
  \text{polarity}(u,t) \;=\;
    \frac{1}{2}\!\left(\frac{P-N}{P+N}+1\right)
    - \frac{1}{2}\!\left(\frac{A-N}{A+N}+1\right),
\end{equation}
where $P$, $N$, and $A$ are the counts of pro-Brexit, neutral, and anti-Brexit comments by $u$ in period $t$.
Polarity lies in $[-1,1]$: $+1$ is strongly pro-Brexit, $-1$ strongly anti-Brexit, and $0$ indicates either exclusively neutral posting or a balanced pro/anti mix (affecting $<$4\% of users).
The metric reflects \emph{revealed preference} (what users actually post) rather than latent conviction: a user who posts one pro-Brexit comment among 99 neutral contributions receives a polarity near zero, correctly weighting the rarity of their partisan signal.

\begin{figure}[tb]
    \centering
    \newcommand\myheight{0.175}
    \subfloat[]{
        \includegraphics[height=\myheight\textheight]{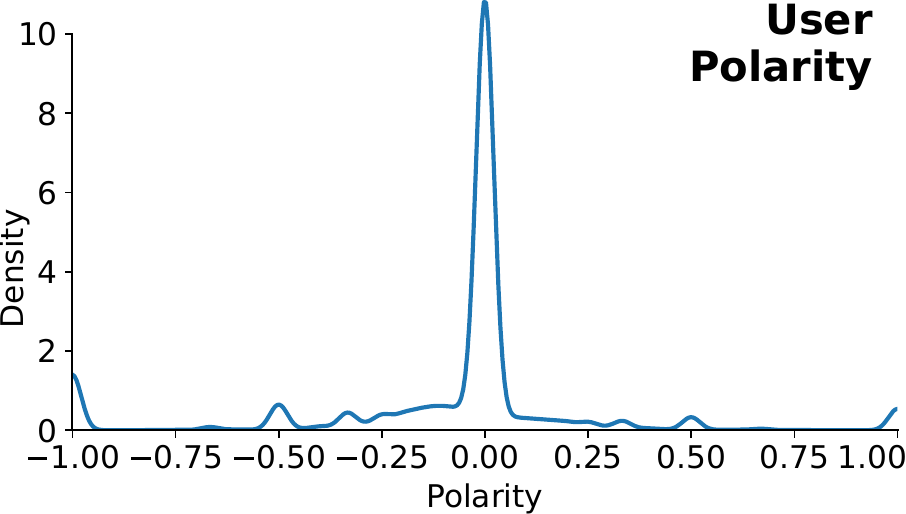}\label{fig:polarity_pdf}
    }\subfloat[]{
        \includegraphics[height=\myheight\textheight]{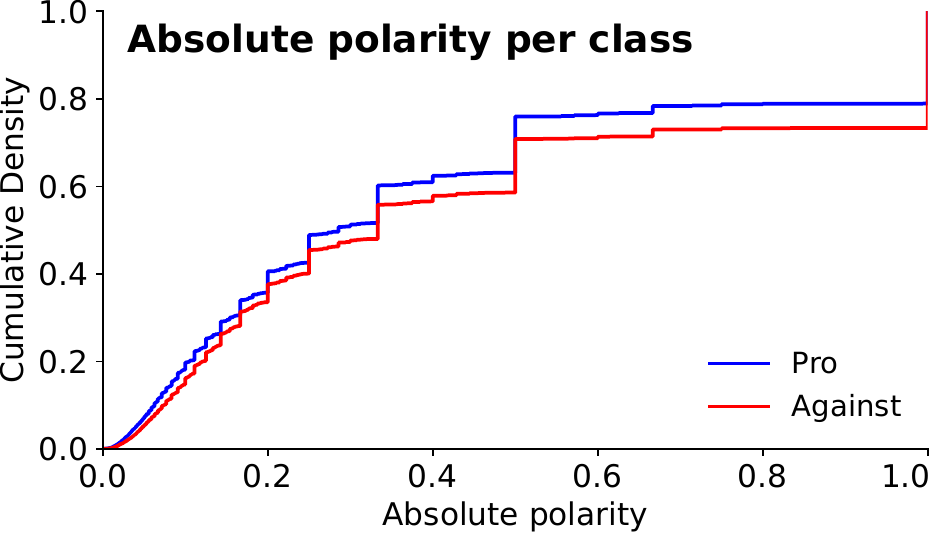}\label{fig:polarity_cdf}
    }\caption{
        \textbf{(a)} PDF of the continuous polarity distribution across all user--period pairs.
        \textbf{(b)} CDF of absolute polarity for pro-leaning ($>0$) and anti-leaning ($<0$) users.
    }
\end{figure}

\Cref{fig:polarity_pdf} reveals that 58\% of user--period pairs have polarity exactly zero, yet the remaining 42\% exhibit a rich spectrum of opinion invisible to discrete labelling.
Anti-Brexit users are slightly more polarised (\cref{fig:polarity_cdf}): the subreddit's anti-Brexit majority attracts more strongly opinionated participants on that side.

\paragraph{Network polarisation features.}
User-level polarity extends to the interaction network.
Let $\vec{n}_i$ denote the set of users that~$i$ has interacted with (replied to or received a reply from) in a given period.
We define two per-edge quantities.
The \emph{interaction polarity} $\text{polarity}(\vec{n}_i)$ is the vector of polarities of~$i$'s interaction partners, capturing the ideological composition of~$i$'s neighbourhood.
The \emph{edge homogeneity} $\text{polarity}(i) \cdot \text{polarity}(\vec{n}_i)$ is the element-wise product of~$i$'s polarity with each partner's polarity: positive entries indicate like-minded interactions, negative entries cross-cutting exchanges (inspired by~\cite{Vicario2016}).
Both are summarised by their mean per user per period and used in \cref{sec:echo} to measure echo chamber structure and predict future polarity.

\section{Polarisation Dynamics Analysis}
\label{sec:analysis}

We organise this section around two questions: who stays on the platform, and what do persistent users believe?
The answers build toward a single finding: it is survivorship bias, not opinion change, that drives the polarisation dynamics visible in the data.

\subsection{Who Stays and Who Leaves?}
\label{sec:presence}

\begin{figure}[t]
  \centering
  \subfloat[\label{fig:presence_patterns}]{\adjustbox{valign=c}{\small\begin{tabular}{lc}
        \toprule
        \textbf{Pattern} & \textbf{Freq.} \\
        \midrule
        $\blacksquare$                                 & 70.5\% \\
        $\blacksquare\blacksquare$                     & 5.0\%  \\
        $\blacksquare\,\square\,\blacksquare$          & 1.3\%  \\
        $\blacksquare\,\square\,\square\,\blacksquare$ & 1.0\%  \\
        $\blacksquare\blacksquare\blacksquare$         & 0.9\%  \\
        \bottomrule
      \end{tabular}}\vphantom{\adjustbox{valign=c}{\includegraphics[width=0.72\columnwidth]{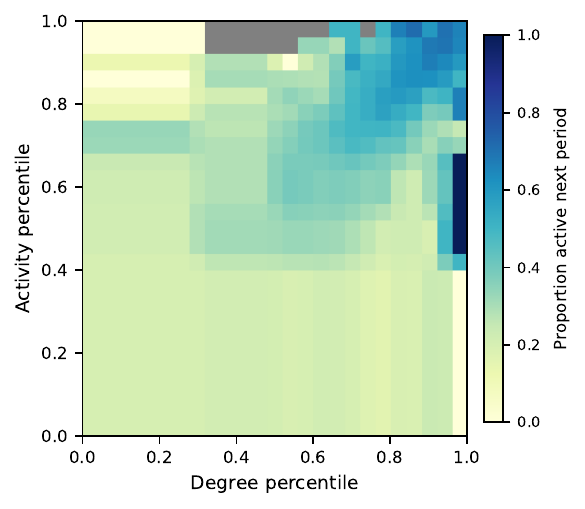}}}}\hfill \subfloat[\label{fig:heat_map}]{\adjustbox{valign=c}{\includegraphics[width=0.72\columnwidth]{new_heat_map_v2}}}\caption{\textbf{(a)} Most frequent user presence patterns ($\blacksquare$~=~present;
    $\square$~=~absent): 70.5\% of users appear in a single period, and fewer than 1\%
    sustain engagement across three or more consecutive periods.
    \textbf{(b)} Proportion of users who remain in the next period, binned by activity
    and degree percentiles. Higher engagement strongly correlates with staying.
    The exception cluster ($>$96th percentile degree, 44th--64th percentile activity)
    contains two highly polarised anti-Brexit broadcasters who accumulate replies without
    posting frequently.}
  \label{fig:stay_leave}
\end{figure}

All figures in this section concern \emph{active contributors} only; lurkers are invisible to our data, so ``returning'' means returning \emph{and} posting again.
The \rbrexit community is dominated by transient participants: 70.5\% of users appear in only one period, and fewer than 1\% sustain engagement across three or more consecutive periods (\cref{fig:presence_patterns}).
Before asking how opinions change, we ask what determines whether an active contributor returns at all.

\paragraph{Activity and degree.}
We define \emph{activity} as the number of comments a user posts per period, and \emph{degree} as the number of unique interaction partners.
\Cref{fig:heat_map} shows the retention heat map across activity and degree percentiles: higher values of both strongly correlate with remaining.
One exception stands out: a cluster above the 96th percentile of degree but only in the 44th--64th percentile of activity: high reach, low output.
Only two users occupy this cluster, both strongly anti-Brexit (polarity $-0.50$ and $-0.667$); one posted across nine periods, favouring short provocative content (e.g., ``Stop the madness. Stop Brexit.''), returning to provoke and \emph{broadcast} their opinions, not to deliberate.

\paragraph{Influential users.}
Interacting with \emph{prolific} users (those at or above the 99th percentile of degree or activity, with 73\% qualifying on both) substantially increases the likelihood of returning: users who interacted with a prolific peer retain at $\approx50\%$, versus $32$--$33\%$ for those who did not ($p \ll 0.001$, $\chi^2$ test).
These hub users act as community anchors, doubling a peripheral user's odds of returning; the long-term community structure is sustained by this small connected core.

\begin{figure}[tb]
  \centering
  \newcommand\myheight{0.175}
  \subfloat[]{
    \includegraphics[height=\myheight\textheight]{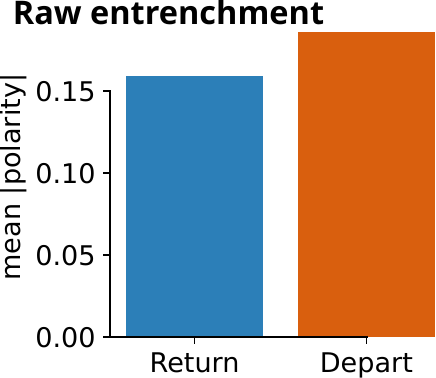}\label{fig:raw-entrenchment}
  }\subfloat[]{
    \includegraphics[height=\myheight\textheight]{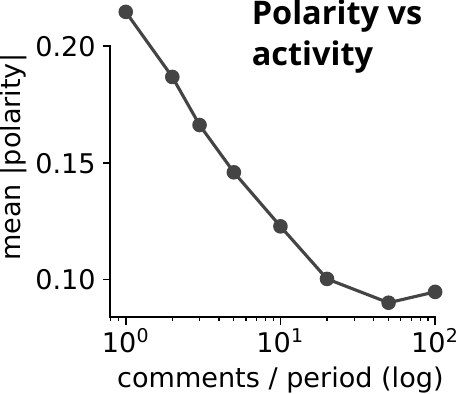}\label{fig:polarity-vs-activity}
  }\subfloat[]{
    \includegraphics[height=\myheight\textheight]{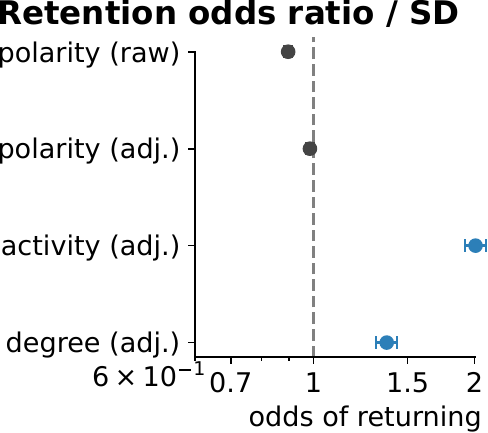}\label{fig:retention-odds-ratio}
  }\caption{
    Retention is selective on engagement, not stance.
    \textbf{(a)}~A raw comparison makes departing users look slightly more polarised.
    \textbf{(b)}~Mean $|\text{polarity}|$ falls with activity: single-post users saturate the metric.
    \textbf{(c)}~Retention odds ratios (per SD, $95\%$~CI): controlling for activity and degree, polarity's effect collapses to~$1$ (no effect), while activity dominates.
  }
  \label{fig:survivor_engagement}
\end{figure}

\begin{figure}[tb]
  \centering
  \newcommand\myheight{0.15}
  \subfloat[]{
    \includegraphics[height=\myheight\textheight]{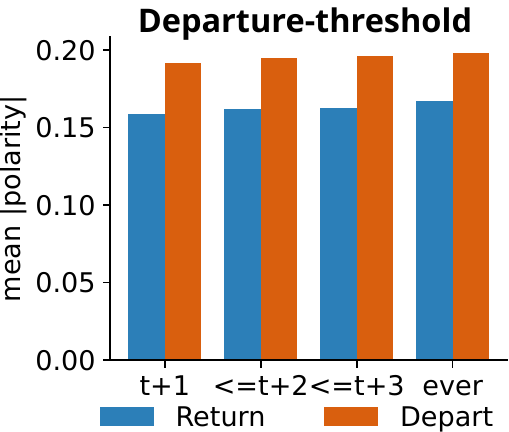}\label{fig:departure-threshold}
  }\subfloat[]{
    \includegraphics[height=\myheight\textheight]{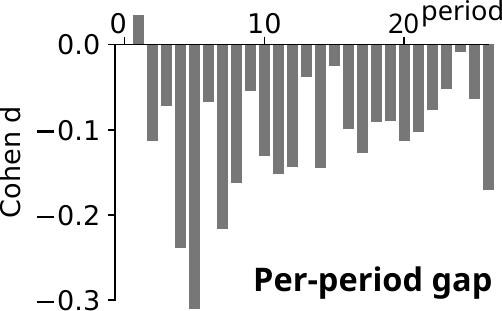}\label{fig:per-period-gap}
  }\subfloat[]{
    \includegraphics[height=\myheight\textheight]{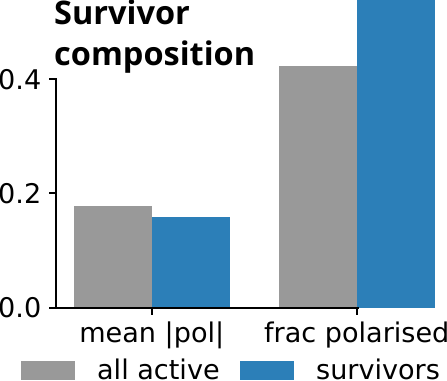}\label{fig:survivor-composition}
  }\caption{
    Robustness of the engagement-driven retention finding, and the resulting selection bias.
    \textbf{(a)}~The raw stayer--leaver entrenchment gap (mean $|\text{polarity}|$) is stable across four definitions of departure (no return at $t{+}1$, within $\leq t{+}2$, $\leq t{+}3$, or ever), holding at Cohen's $d\approx-0.11$ throughout.
    \textbf{(b)}~Per-period Cohen's $d$ (stayer~$-$~leaver): leavers are at least as polarised as stayers in $25$ of $26$ testable periods, so the gap is not an aggregation artefact.
    \textbf{(c)}~Conditioning on retention yields a non-representative survivor subpopulation: relative to all active users, survivors are more often non-neutral ($54\%$ vs.\ $42\%$) yet milder in magnitude (mean $|\text{polarity}|$ $0.157$ vs.\ $0.177$).}
  \label{fig:survivor_robust}
\end{figure}

\paragraph{Engagement, not stance, decides who returns.}
A raw comparison makes departing users look slightly more polarised (\cref{fig:raw-entrenchment}), but this is a measurement artefact.
The polarity metric saturates at low activity (\cref{fig:polarity-vs-activity}), e.g. one partisan comment scores $|\text{polarity}|{=}1$; 
one-off posters (the bulk of departures: $76\%$ of single-post authors never return) dominate it.
Controlling for activity and degree, polarity carries no significant association with retention (odds ratio $0.98$/SD, $95\%$~CI $[0.97,1.00]$, $p=0.10$; \cref{fig:retention-odds-ratio}), whereas a one-SD rise in activity doubles the odds of returning.
The most strident voices (low-activity, maximally polarised single posts) are thus disproportionately transient (returners average $18.5$ versus $3.8$ comments per period).
This result holds however we define departure.
Whether a user counts as departed when they skip the next period, the next two or three periods, or never return at all, the gap between stayers and leavers barely moves: leavers stay slightly more polarised throughout, by Cohen's $d\approx-0.11$ (\cref{fig:departure-threshold}).
The strict next-period cutoff does not drive this, since $28\%$ of the users it marks as departed reappear in a later period.
The gap also holds period by period rather than only in aggregate: leavers are at least as polarised as stayers in $25$ of the $26$ testable periods (\cref{fig:per-period-gap}).
Conditioning on retention still leaves a survivor minority that looks little like the wider population (\cref{fig:survivor-composition}): compared with all active users ($n{=}56{,}111$), survivors ($n{=}23{,}540$) take a side more often ($54\%$ vs.\ $42\%$) but hold it less strongly (mean $|\text{polarity}|$ $0.157$ vs.\ $0.177$), because users who comment enough to stay register a lean yet rarely push the metric to its extremes.
Longitudinal stance analyses therefore rest on a self-selected, highly active minority rather than a representative sample of opinion.

\begin{figure}[tb]
    \centering
    \subfloat[]{\includegraphics[height=0.163\textheight]{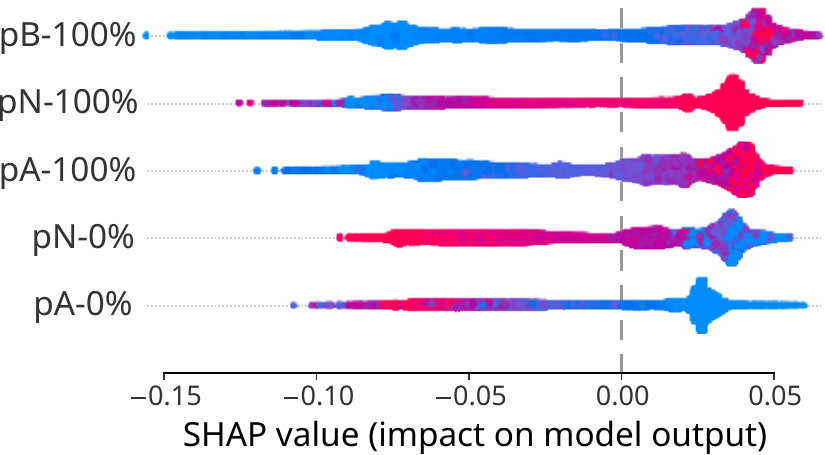}
        \label{fig:shap_presence}}\subfloat[]{\includegraphics[height=0.163\textheight]{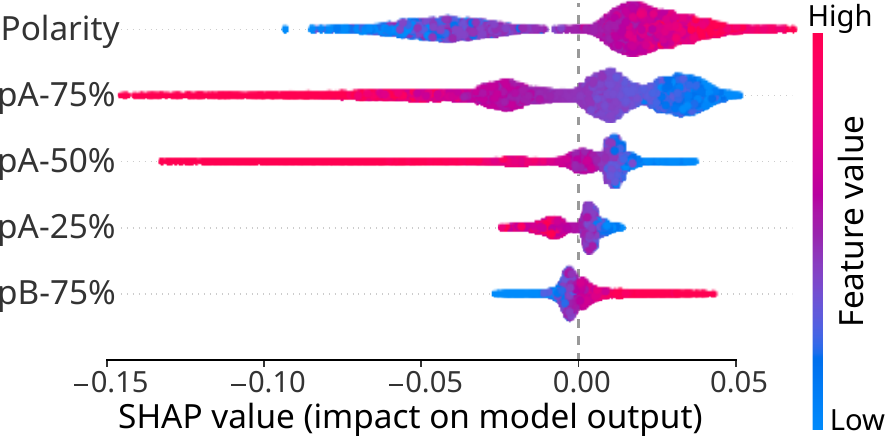}\label{fig:shap_regression}}\caption{\textbf{(a)} SHAP values for future presence: participating in polarised discussions
        (red at \texttt{pB-100\%}) strongly predicts remaining.
        \textbf{(b)} SHAP values for future polarity regression (F0123): high current polarity
        (red) strongly predicts high future polarity.} 
\end{figure}

\paragraph{SHAP analysis of future presence.}
We train a random forest classifier for future presence (remain vs.\ leave) and apply SHAP analysis (\cref{fig:shap_presence}).
Recall that $p_s\text{-}y\%$ denotes the $y$th percentile of stance-$s$ saturation across a user's discussions, where $s \in \{\text{B (pro-Brexit)}, \text{A (anti-Brexit)}, \text{N (neutral)}\}$ (full inventory in Online Appendix~D~\cite{appendix}).
Participating in at least one highly polarised discussion (\texttt{pB-100\%}, \texttt{pA-100\%}) strongly predicts remaining.
Conversely, engaging \emph{only} in maximally anti-Brexit discussions (\texttt{pA-0\%} high) or only in neutral threads (\texttt{pN-0\%} high) \emph{decreases} the likelihood of remaining.
Broad participation across polarised discussions, rather than exclusive immersion in any single extreme, characterises users who return.

Across all four analyses, the pattern is consistent: retention is driven by engagement depth and breadth, not by ideological content alone.

\subsection{Persistent Users Live in Echo Chambers}
\label{sec:echo}

\begin{figure}[tb]
    \centering
      \newcommand\myheight{0.195}
    \subfloat[]{\includegraphics[height=\myheight\textheight]{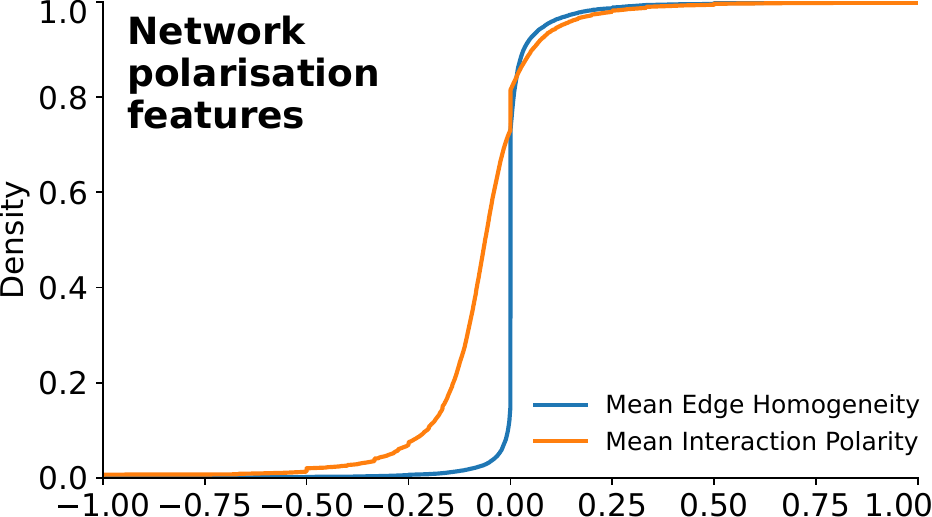}\label{fig:echo_cdf}}\subfloat[]{\includegraphics[height=\myheight\textheight]{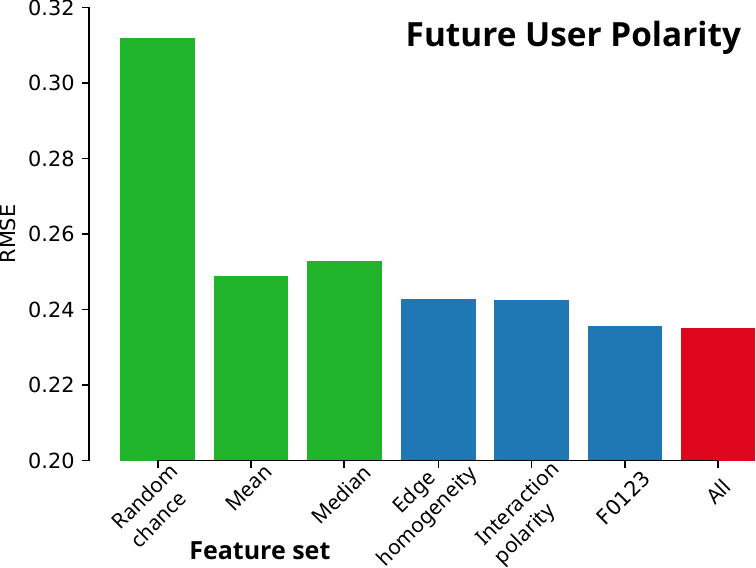}\label{fig:regression_res}}\caption{\textbf{(a)} CDF of mean interaction polarity and mean edge homogeneity per user.
        The rightward skew of edge homogeneity confirms that like-minded interactions dominate.
        \textbf{(b)} Random forest regression RMSE for future polarity prediction (lower is better).
        Green bars are random baselines; blue bars are the F0123 feature set (prior work features augmented with features constructed here); red bars use all features including edge polarity.
        Edge polarity features do not improve significantly over the F0123 descriptor set.}
\end{figure}

\paragraph{Echo chamber evidence.}
Edge homogeneity (\cref{sec:polarity}) quantifies ideological alignment: \cref{fig:echo_cdf} shows that nearly 40\% of user-level interactions occur between like-minded individuals, while cross-cutting exchanges are rare.
Anti-Brexit users are slightly more polarised (consistent with \cref{sec:polarity}).

\paragraph{SHAP analysis of future polarity.}
F0123 comprises four feature groups~\cite{Largeron2021}: textual content (F0), user activity statistics (F1, F2), and discussion-composition descriptors (F3); the full inventory is in Online Appendix~D~\cite{appendix}.
\Cref{fig:regression_res} compares RMSE across feature sets: all outperform the trivial baselines, and adding edge polarity features to F0123 yields negligible improvement, because F0123 already encodes network-level information through its discussion-composition descriptors ($p_s\text{-}y\%$; Online Appendix~D~\cite{appendix}).
SHAP analysis~\cite{lundberg2017unified} (\cref{fig:shap_regression}) reveals that current polarity is the dominant predictor, with the anti-Brexit sentiment level of a user's discussions (\texttt{pA-75\%}, \texttt{pA-50\%}) as the next strongest signal.
This is consistent with echo-chamber effects: immersion in predominantly anti-Brexit discussions correlates with maintaining or deepening an anti-Brexit stance, rather than with cross-cutting opinion change.

The persistent users who populate longitudinal analyses interact largely within like-minded echo chambers, and their polarity changes little over time.

\section{Discussion and Conclusion}
\label{sec:discussion}

We presented a framework for measuring and analysing polarisation dynamics on structured discussion platforms, demonstrated on the \rbrexit subreddit spanning the full Brexit saga (2015--2021).
Our three contributions (the \rbrexit dataset, a continuous polarity pipeline, and behavioural analysis) answer a question the literature has long asked: \emph{how does exposure to opposing views change opinions?}
On \rbrexit, it mostly does not: the persistent core sits in echo chambers and barely moves, while the wider population churns; who remains is selected by engagement, not conviction.
Future work includes multi-subreddit and multi-topic analysis, causal modelling of the participation--stance feedback loop~\cite{Tian2026Causal}, and whether next-generation classifiers alter the magnitude (though, given the survivorship mechanism, not the direction) of these findings.

\textbf{Richer models do not improve prediction.}
We augmented F0123 with three progressively more expressive architectures: \emph{stance triads} (triadic-closure counts labelled by node stance), a \emph{Graph Attention Network} (GAT) over the per-period interaction graph, and a bidirectional LSTM over the comment sequence.
Across all variants, test F1 moves at most from $0.368$ to $0.371$ (full results in Online Appendix~E~\cite{appendix}).
This consistent null is structural: once discussion-composition features encode which types of discussions a user selects into, finer interaction topology carries no additional signal: self-selection sets a predictive ceiling.

\textbf{Platform-specific challenges for NLP.} 
We tested augmenting the training data with weakly-labelled texts identified via stance-associated keywords (e.g., ``remoaner'' for pro-Brexit, ``Brexshit'' for anti-Brexit).
The augmented BERT-Reddit model degrades performance (F1~$= 0.46$ vs.\ $0.555$ unaugmented), aggressively and near-randomly predicting the pro-Brexit class.
The root cause is that Reddit users routinely quote opponents: pro-Brexit keywords appear in anti-Brexit posts that mock or rebut them.
This quoting behaviour systematically corrupts keyword augmentation; NLP pipelines must account for embedded adversarial quotations.

\textbf{Limitations.}
Crucially, the central survivorship finding (70.5\% of users appear in a single period, and departure tracks engagement rather than ideology) rests on binary presence indicators, not classifier labels, so classifier error cannot affect it.
Polarity-based analyses (echo chambers, future-polarity regression) do carry classifier error, but rest on the direction and rank-ordering of effects, not precise counts; more accurate labels would strengthen, not weaken, the survivorship mechanism.
The regression and SHAP analyses are predictive and correlational; they identify features that co-vary with future polarity, not causes of opinion change.
The continuous polarity metric assigns zero both to genuinely neutral users and to the small fraction ($<$4\%) who post equal numbers of pro- and anti-Brexit comments, a conflation resolvable by treating stance counts as simplex coordinates.
Reddit's API surfaces only content, not reads: lurkers are invisible, and apparent ``departure'' may reflect passive participation rather than genuine disengagement.
We define departure by posting absence; although a minority of single-period absences are temporary returns, the retention patterns are stable under stricter, multi-period definitions (\cref{fig:survivor_robust}).
Event-based period boundaries are motivated by the substantive goal of capturing coherent phases of the Brexit debate; coarser temporal windows would lower the 70.5\% transience figure, making the claim more conservative, not more extreme.
Scope is limited to a single subreddit on a single topic; the mechanism we identify is unlikely to be platform-specific (see \cref{sec:related} and the discussion of generalisability below).

\textbf{Broader implications.}
For \emph{platform designers}: algorithmic interventions that target the most active users (those most exposed to any recommendation or moderation signal) will be ineffective if those users are already entrenched, and may even provoke community resistance and migration to less-moderated platforms~\cite{Johns2024}; interventions should instead prioritise new and peripheral users, who are at greatest risk of early departure, reaching them before they leave.
Nudges at first engagement (e.g., surfacing cross-cutting content to newcomers) are more likely to shift trajectories than interventions on long-term participants whose stance has stabilised; intervention opportunities are greatest at the early stages of the radicalisation pathway~\cite{Booth2024}, and style-tailored messaging can reach otherwise-resistant fringe communities before their positions harden~\cite{Lee2024bubbles}.
Operationally, the pipeline (in-domain stance classification, continuous polarity, and joint presence--stance analysis) is deployable as a community-monitoring tool that flags when an apparent shift in opinion is in fact a shift in \emph{who} is posting.
For \emph{researchers}: ``future stance prediction'' models are structurally misleading unless they account for the selection bias of user retention.
We advocate jointly modelling presence and stance via a survival-regression architecture in which departure and future polarity share latent user representations, rather than treating retention as a fixed precondition for stance analysis.
The mechanism is unlikely to be unique to \rbrexit: Oswald et al.~\cite{oswald2025disentangling} show experimentally that online political discussions are dominated by a systematically atypical active minority, supporting self-selection as a confound beyond any single platform.

\begin{credits}

\subsubsection{Acknowledgements}
This research was supported by the Advanced Strategic Capabilities Accelerator (ASCA), the Defence Science and Technology Group, the Defence Innovation Network, and the Australian Academy of Science.

\subsubsection{Ethics Statement}
Data were collected from the publicly accessible Reddit platform via the Pushshift API~\cite{pushshift}; no personally identifiable information beyond posted usernames was retained.
The study is observational analysis of publicly available content.
\end{credits}

\bibliographystyle{splncs04}

\begin{thebibliography}{10}
\providecommand{\url}[1]{\texttt{#1}}
\providecommand{\urlprefix}{URL }
\providecommand{\doi}[1]{https://doi.org/#1}

\bibitem{appendix}
Online appendix: {Online Polarisation Dynamics on Reddit}. \url{https://arxiv.org/pdf/2603.07509#page=19} (2026)

\bibitem{aldayel2021stance}
AlDayel, A., Magdy, W.: Stance detection on social media: State of the art and trends. Inf. Process. Manag.  \textbf{58}(4),  102597 (2021)

\bibitem{alturayeif2023systematic}
Alturayeif, N., Luqman, H., Ahmed, M.: A systematic review of machine learning techniques for stance detection and its applications. Neural Comput. Appl.  \textbf{35},  5113--5144 (2023). \doi{10.1007/s00521-023-08285-7}

\bibitem{mturkapi}
{Amazon Mechanical Turk}: Amazon mechanical turk. \url{https://www.mturk.com/}, accessed: 2021

\bibitem{Booth2024}
Booth, E., Lee, J., Rizoiu, M.A., Farid, H.: Conspiracy, misinformation, radicalisation: Understanding the online pathway to indoctrination and opportunities for intervention. J. Sociol.  \textbf{60}(2),  440--457 (2024). \doi{10.1177/14407833241231756}

\bibitem{bruns2017echo}
Bruns, A.: Echo chamber? what echo chamber? reviewing the evidence. In: Future of Journalism Conf. (FOJ17) (2017)

\bibitem{Calderon2024a}
Calderon, P., Ram, R., Rizoiu, M.A.: Opinion market model: Stemming far-right opinion spread using positive interventions. In: Proc. ICWSM. vol.~18, pp. 177--190 (2024). \doi{10.1609/icwsm.v18i1.31306}

\bibitem{Calderon2024b}
Calderon, P., Rizoiu, M.A.: What drives online popularity: Author, content or sharers? {Estimating} spread dynamics with {Bayesian} mixture {Hawkes}. In: Proc. ECML-PKDD. pp. 142--160 (2024). \doi{10.1007/978-3-031-70362-1_9}

\bibitem{cinus2022recommenders}
Cinus, F., Minici, M., Monti, C., Bonchi, F.: The effect of people recommenders on echo chambers and polarization. In: Proc. ICWSM. vol.~16, pp. 90--101 (2022). \doi{10.1609/icwsm.v16i1.19275}

\bibitem{colacrai2024compass}
Colacrai, E., Cinus, F., {De Francisci Morales}, G., Starnini, M.: Navigating multidimensional ideologies with {Reddit}'s political compass: Economic conflict and social affinity. In: Proc. WWW. pp. 2582--2593 (2024). \doi{10.1145/3589334.3645606}

\bibitem{Colleoni:2014}
Colleoni, E., Rozza, A., Arvidsson, A.: Echo chamber or public sphere? predicting political orientation and measuring political homophily in {Twitter} using big data. J. Commun.  \textbf{64}(2),  317--332 (2014). \doi{10.1111/jcom.12084}

\bibitem{conover2011political}
Conover, M.D., Ratkiewicz, J., Francisco, M., Gon{\c{c}}alves, B., Menczer, F., Flammini, A.: Political polarization on {Twitter}. In: Proc. ICWSM (2011)

\bibitem{Vicario2016}
Del~Vicario, M., Bessi, A., Zollo, F., Petroni, F., Scala, A., Caldarelli, G., Stanley, H.E., Quattrociocchi, W.: The spreading of misinformation online. PNAS  \textbf{113}(3),  554--559 (2016). \doi{10.1073/pnas.1517441113}

\bibitem{devlin2018bert}
Devlin, J., Chang, M.W., Lee, K., Toutanova, K.: Bert: Pre-training of deep bidirectional transformers for language understanding. arXiv preprint arXiv:1810.04805  (2018)

\bibitem{efstratiou2024deliberate}
Efstratiou, A.: Deliberate exposure to opposing views and its association with behavior and rewards on political communities. In: Proc. WWW. pp. 2347--2358 (2024). \doi{10.1145/3589334.3645375}

\bibitem{efstratiou2023nonpolar}
Efstratiou, A., Blackburn, J., Caulfield, T., Stringhini, G., Zannettou, S., {De Cristofaro}, E.: Non-polar opposites: Analyzing the relationship between echo chambers and hostile intergroup interactions on {Reddit}. In: Proc. ICWSM. vol.~17, pp. 197--208 (2023). \doi{10.1609/icwsm.v17i1.22138}

\bibitem{gatto2023cot}
Gatto, J., Sharif, O., Preum, S.M.: Chain-of-thought embeddings for stance detection on social media. In: Findings of EMNLP. pp. 4154--4161 (2023). \doi{10.18653/v1/2023.findings-emnlp.273}

\bibitem{gera2025deep}
Gera, P., Neal, T.J.: Deep learning in stance detection: A survey. ACM Comput. Surv.  \textbf{58}(1),  1--37 (2025). \doi{10.1145/3744641}

\bibitem{hofmann2022politosphere}
Hofmann, V., Sch{\"u}tze, H., Pierrehumbert, J.B.: The {Reddit} politosphere: A large-scale text and network resource of online political discourse. In: Proc. ICWSM. vol.~16, pp. 1259--1267 (2022). \doi{10.1609/icwsm.v16i1.19377}

\bibitem{Johns2024}
Johns, A., Bailo, F., Booth, E., Rizoiu, M.A.: Labelling, shadow bans and community resistance: Did {Meta}'s strategy to suppress rather than remove {COVID} misinformation and conspiracy theory on {Facebook} slow the spread? Media Int. Aust.  (2024). \doi{10.1177/1329878X241236984}

\bibitem{Kong2022}
Kong, Q., Booth, E., Bailo, F., Johns, A., Rizoiu, M.A.: Slipping to the extreme: A mixed method to explain how extreme opinions infiltrate online discussions. In: Proc. ICWSM. vol.~16, pp. 524--535 (2022). \doi{10.1609/icwsm.v16i1.19312}

\bibitem{Largeron2021}
Largeron, C., Mardale, A., Rizoiu, M.A.: Linking the dynamics of user stance to the structure of online discussions. In: Proc. IDA, pp. 275--286 (2021). \doi{10.1007/978-3-030-74251-5_22}

\bibitem{Lee2024bubbles}
Lee, J., Booth, E., Farid, H., Rizoiu, M.A.: Popping misinformation bubbles: Results from style-based messaging into online misinformation ecosystems. Research Square preprint (2024). \doi{10.21203/rs.3.rs-4528128/v2}

\bibitem{Lee2025facts}
Lee, J., Booth, E., Farid, H., Rizoiu, M.A.: Misinformation is not about bad facts: An analysis of the production and consumption of fringe content. {EPJ} Data Science  \textbf{14}(1), ~50 (2025). \doi{10.1140/epjds/s13688-025-00567-5}

\bibitem{Lee2026}
Lee, J., Tian, L., Brillantes, A., Mihăiţă, A.S., Rizoiu, M.A.: Long live fine-tuning: Task-specific transformers outperform zero-shot {LLMs} for misinformation response classification on {Reddit}  (6 2026). \doi{10.48550/arXiv.2606.04274}

\bibitem{lenti2025variational}
Lenti, J., Silvestri, F., {De Francisci Morales}, G.: Variational inference of parameters in opinion dynamics models. In: Proc. ICWSM. vol.~19, pp. 2622--2627 (2025). \doi{10.1609/icwsm.v19i1.35963}

\bibitem{lorge2024stentconv}
Lorge, I., Zhang, L., Dong, X., Pierrehumbert, J.: {STEntConv}: Predicting disagreement between {Reddit} users with stance detection and a signed graph convolutional network. In: Proc. LREC-COLING. pp. 15273--15284 (2024)

\bibitem{lundberg2017unified}
Lundberg, S.M., Lee, S.I.: A unified approach to interpreting model predictions. In: Proc. NeurIPS. pp. 4768--4777 (2017)

\bibitem{mohammad2016semeval}
Mohammad, S., Kiritchenko, S., Sobhani, P., Zhu, X., Cherry, C.: Semeval-2016 task 6: Detecting stance in tweets. In: Proc. SemEval. pp. 31--41 (2016)

\bibitem{mok2023echo}
Mok, L., Inzlicht, M., Anderson, A.: Echo tunnels: Polarized news sharing online runs narrow but deep. In: Proc. ICWSM. vol.~17, pp. 662--673 (2023)

\bibitem{monti2023demographic}
Monti, C., D'Ignazi, J., Starnini, M., {De Francisci Morales}, G.: Evidence of demographic rather than ideological segregation in news discussion on {Reddit}. In: Proc. WWW. pp. 2777--2786 (2023)

\bibitem{nguyen2020bertweet}
Nguyen, D.Q., Vu, T., Nguyen, A.T.: {BERTweet}: A pre-trained language model for {English} tweets. arXiv preprint arXiv:2005.10200  (2020)

\bibitem{okawa2022sinn}
Okawa, M., Iwata, T.: Predicting opinion dynamics via sociologically-informed neural networks. In: Proc. KDD. pp. 1306--1316 (2022). \doi{10.1145/3534678.3539228}

\bibitem{oswald2025disentangling}
Oswald, L., Schulz, W.S., Lorenz-Spreen, P.: Disentangling participation in online political discussions with a collective field experiment. Sci. Adv.  \textbf{11}(50),  eady8022 (2025). \doi{10.1126/sciadv.ady8022}

\bibitem{petruzzellis2023opinion}
Petruzzellis, F., Bonchi, F., {De Francisci Morales}, G., Monti, C.: On the relation between opinion change and information consumption on {Reddit}. In: Proc. ICWSM. vol.~17, pp. 710--719 (2023)

\bibitem{Ram2024}
Ram, R., Rizoiu, M.A.: Empirically measuring online social influence. {EPJ} Data Science  \textbf{13}(1), ~53 (2024)

\bibitem{Ram2026}
Ram, R., Rizoiu, M.A.: Conductance and influence-capital: Modeling online social influence. {EPJ} Data Science  \textbf{15}(1), ~56 (2026). \doi{10.1140/epjds/s13688-026-00650-5}

\bibitem{Ram2025}
Ram, R., Thomas, E., Kernot, D., Rizoiu, M.A.: Practical guidelines for ideology detection pipelines and psychosocial applications. In: Proc. ICWSM. vol.~19, pp. 1630--1648 (2025)

\bibitem{pushshift}
{/r/datasets mod team}: Pushshift {Reddit} {API}. \url{https://pushshift.io/} (2019), accessed: 2021

\bibitem{schick2021generating}
Schick, T., Sch{\"u}tze, H.: Generating datasets with pretrained language models. arXiv preprint arXiv:2104.07540  (2021)

\bibitem{Schneider2023}
Schneider, P.J., Rizoiu, M.A.: The effectiveness of moderating harmful online content. PNAS  \textbf{120}(34),  e2307360120 (2023). \doi{10.1073/pnas.2307360120}

\bibitem{Schneider2025}
Schneider, P.J., Tian, L., Rizoiu, M.A.: Learning to make friends: Coaching {LLM} agents toward emergent social ties. In: Scaling Environments for Agents (SEA) @ NeurIPS 2025 (2025)

\bibitem{steel2024multi}
Steel, B., Ruths, D.: Multi-target user stance discovery on {Reddit}. In: Proc. WASSA. pp. 200--214 (2024)

\bibitem{sun2019fine}
Sun, C., Qiu, X., Xu, Y., Huang, X.: How to fine-tune {BERT} for text classification? In: Proc. CCL. pp. 194--206. Springer (2019)

\bibitem{Tian2025ICMamba}
Tian, L., Booth, E., Bailo, F., Droogan, J., Rizoiu, M.A.: Before it's too late: A state space model for the early prediction of misinformation and disinformation engagement. In: Proc. WWW. pp. 5244--5254 (2025)

\bibitem{Tian2026}
Tian, L., Rizoiu, M.A.: {DREAMS}: A social exchange theory-informed modeling of misinformation engagement on social media. In: Proc. WWW (2026). \doi{10.1145/3774904.3792681}

\bibitem{Tian2026Causal}
Tian, L., Rizoiu, M.A.: Estimating online influence needs causal modeling! {Counterfactual} analysis of misinformation engagement on social media. In: Proc. AAAI. vol.~40, pp. 1078--1086 (2026). \doi{10.1609/aaai.v40i2.37078}

\bibitem{Tian2025XTroll}
Tian, L., Zhang, X., Kim, M.M.H., Biggs, J., Rizoiu, M.A.: {X-Troll}: {eXplainable} detection of state-sponsored information operations agents. In: Proc. CIKM. pp. 2874--2884 (2025). \doi{10.1145/3746252.3761028}

\bibitem{weinzierl2024tree}
Weinzierl, M., Harabagiu, S.: Tree-of-counterfactual prompting for zero-shot stance detection. In: Proc. ACL. pp. 861--880 (2024)

\bibitem{wojcieszak2009online}
Wojcieszak, M.E., Mutz, D.C.: Online groups and political discourse: Do online discussion spaces facilitate exposure to political disagreement? J. Commun.  \textbf{59}(1),  40--56 (2009)

\bibitem{Yuan2024}
Yuan, L., Rizoiu, M.A.: Generalizing hate speech detection using multi-task learning: A case study of political public figures. Comput. Speech Lang.  \textbf{89},  101690 (2025). \doi{10.1016/j.csl.2024.101690}

\bibitem{Yuan2025}
Yuan, L., Schneider, P.J., Rizoiu, M.A.: Behavioral homophily in social media via inverse reinforcement learning: A reddit case study. In: Proc. WWW. pp. 576--589 (2025)

\bibitem{Yuan2023}
Yuan, L., Wang, T., Ferraro, G., Suominen, H., Rizoiu, M.A.: Transfer learning for hate speech detection in social media. J. Comput. Soc. Sci.  \textbf{6}(2),  1081--1101 (2023). \doi{10.1007/s42001-023-00224-9}

\bibitem{ziems2024llms}
Ziems, C., Held, W., Shaikh, O., Chen, J., Zhang, Z., Yang, D.: Can large language models transform computational social science? Comput. Linguist.  \textbf{50}(1),  237--291 (2024). \doi{10.1162/coli_a_00502}

\end{thebibliography}

\begin{thebibliography}{1}
\providecommand{\url}[1]{\texttt{#1}}
\providecommand{\urlprefix}{URL }
\providecommand{\doi}[1]{https://doi.org/#1}

\bibitem{callison2010creating}
Callison-Burch, C., Dredze, M.: Creating speech and language data with {Amazon}'s {Mechanical Turk}. In: Proc. NAACL-HLT Workshop. pp. 1--12 (2010)

\bibitem{gallicchio2017randomized}
Gallicchio, C., Mart{\'\i}n-Guerrero, J.D., Micheli, A., Soria-Olivas, E.: Randomized machine learning approaches: Recent developments and challenges. In: ESANN (2017)

\bibitem{hochreiter1997long}
Hochreiter, S., Schmidhuber, J.: Long short-term memory. Neural Comput.  \textbf{9}(8),  1735--1780 (1997). \doi{10.1162/neco.1997.9.8.1735}

\bibitem{turkerview}
TurkerView: Turkerview. \url{https://turkerview.com/} (2021), accessed: 2021

\bibitem{turkopticon}
Turkopticon. \url{https://turkopticon.net/}, accessed: 2021

\end{thebibliography}

\newpage
\section*{Appendices}

\appendix

This appendix provides supporting material for completeness; it is not required to follow the main paper or reproduce its results.

\section{Brexit Periods: Full Table}
\label{app:periods}

\Cref{tab:period_desc} lists all 27 time periods into which we segment the \rbrexit corpus.
Each period begins on the date of a high-profile political event and runs until the start of the next period (or 28~February~2021 for the final period).
The \emph{\#posts} column counts the total number of Reddit submissions (posts and comments) within the period.

\begin{table}[tbp]
  \caption{The 27 time periods of the \rbrexit dataset, each demarcated by a salient political event in the Brexit process.}
  \label{tab:period_desc}
  \centering
  \small
  \setlength{\tabcolsep}{4pt}
  \begin{tabular}{clrp{7.5cm}}
    \toprule
    \textbf{Per.} & \textbf{Start Date} & \textbf{\#posts} & \textbf{Important Event(s)} \\
    \midrule
    1  & 16/11/2015 &   3,367 & Referendum called; David Cameron resigns \\
    2  & 26/06/2016 &   6,265 & Theresa May accepts Queen's invitation to form government \\
    3  & 14/07/2016 &   3,084 & UK House of Commons votes in favour of Article 50 \\
    4  & 08/12/2016 &   1,466 & Brexit is initiated \\
    5  & 27/01/2017 &   2,300 & Two-year process begins \\
    6  & 30/03/2017 &   4,102 & Brexit negotiations start \\
    7  & 20/06/2017 &  54,505 & White paper finalised; Secretary of State resigns \\
    8  & 09/07/2018 &  23,067 & EU rejects white paper \\
    9  & 22/09/2018 &  15,385 & Brexit withdrawal agreement published \\
    10 & 16/11/2018 &   3,718 & Other 27 EU member states endorse withdrawal agreement \\
    11 & 26/11/2018 &  25,568 & UK Government defeated in withdrawal vote \\
    12 & 16/01/2019 &  54,850 & Second withdrawal vote defeated; extension vote passed \\
    13 & 15/03/2019 &   9,119 & First request for Article 50 extension \\
    14 & 22/03/2019 &  13,414 & Third defeat of UK Government \\
    15 & 30/03/2019 &   9,509 & Second request for Article 50 extension \\
    16 & 25/05/2019 &  27,781 & UK European Parliament elections; Theresa May resigns \\
    17 & 29/08/2019 &  78,434 & Boris Johnson becomes Prime Minister \\
    18 & 10/09/2019 &  19,662 & MPs reject motion to call general election \\
    19 & 25/09/2019 &  18,872 & Supreme Court rules prorogation unlawful \\
    20 & 03/10/2019 &  10,608 & White paper outlining a plan to replace Irish backstop \\
    21 & 18/10/2019 &  17,174 & UK and European Commission revise withdrawal agreement \\
    22 & 30/10/2019 &  13,546 & Third extension of Brexit deadline \\
    23 & 14/12/2019 &  30,510 & Conservatives win general election \\
    24 & 23/01/2020 &  44,958 & Withdrawal Agreement Bill passes parliament \\
    25 & 01/02/2020 &  12,368 & UK begins withdrawal from EU \\
    26 & 18/03/2020 &  52,131 & EU draft proposal for new partnership with UK \\
    27 & 01/01/2021 & 215,382 & UK completes separation from EU \\
    \cmidrule(lr){1-4}
       & 28/02/2021 &         & \textit{dataset ends} \\
    \bottomrule
  \end{tabular}
\end{table}

\section{MTurk Annotation Pipeline: Full Details}
\label{app:mturk}

This appendix provides full details of the crowdsourced annotation pipeline summarised in Section~3.2 of the main paper.
We first describe the task design and HIT interface in \cref{app:mturk:design}.
We then detail the three-step quality control process: parameter optimisation (\cref{app:mturk:step1}), malicious worker detection (\cref{app:mturk:step2}), and the low-confidence filter (\cref{app:mturk:step3}).
Finally, we present the full parameter tuning batch table in \cref{app:mturk:table}.

\subsection{Task Design and HIT Interface} \label{app:mturk:design}

\begin{figure}[h]
  \centering
  \includegraphics[width=\textwidth]{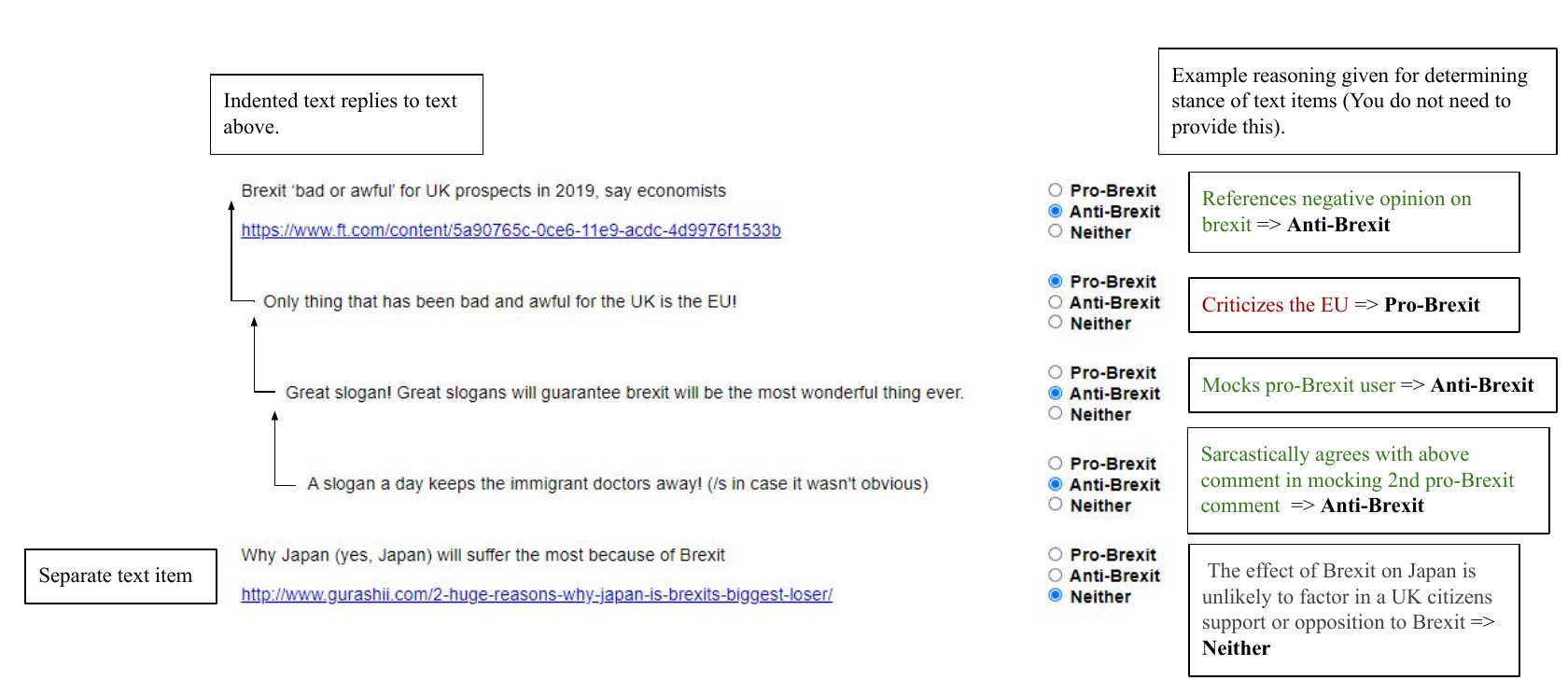}
  \caption{The MTurk annotation interface shown to workers (\emph{Fig.~23 of the IEEE draft}).
  The panel presents a Reddit discussion thread and asks workers to assign one of three stance labels (\emph{pro-Brexit}, \emph{anti-Brexit}, or \emph{neither}) to five submissions per HIT.
  Annotated arrows and text boxes are overlaid for illustration; they do not appear in the actual questionnaire.}
  \label{fig:hit_instr}
\end{figure}

We use Amazon Mechanical Turk (MTurk)~\cite{mturkapi} to gather ground-truth stance labels for a stratified sample of \rbrexit submissions.
Each Human Intelligence Task (HIT) presents workers with a portion of a Reddit discussion thread---including the submission being annotated together with its surrounding reply context---and asks them to label the stance of five submissions per HIT.
To construct each HIT, we take the comment thread of each selected submission (the original post plus its first- and second-level replies), build sets of 2, 3, 4, and 5 comment-long threads, and arrange them into HITs containing exactly five Reddit submissions.
The HIT interface is implemented in bespoke HTML and JavaScript; \cref{fig:hit_instr} shows the annotation panel, which displays the discussion thread and provides instructional tips explaining how workers should interpret the interface and assign a stance label.
Showing the thread context enables workers to use conversational cues (replies, rebuttals, endorsements) to infer stance, and allows them to label adjacent submissions within the same thread simultaneously.
Each submission is labelled by eight workers; the majority label is retained.
We follow Mohammad et al.~\cite{mohammad2016semeval} in adopting three stance classes: \emph{pro-Brexit}, \emph{anti-Brexit}, and \emph{neither}.
A neutral class is deliberately omitted: annotating neutrality is known to be unreliable in polarised settings, where neutral users rarely signal their position explicitly and the absence of pro/anti signals cannot be used to infer neutrality~\cite{mohammad2016semeval}.

\paragraph{Payment.}
To determine a fair reward, we manually annotate a sample of Reddit submissions and measure the average time to complete a single HIT (five submissions) at approximately 45~seconds.
Based on the US federal minimum wage of \$7.25 per hour, we set the per-HIT reward at \$0.10, equivalent to approximately \$8 per hour.
MTurk charges a flat 20\% commission on all rewards; with eight annotators per HIT, the total net cost is \$0.96 per HIT, covering all eight labels for the five included submissions.
To ensure our payment is considered fair by the worker community, we monitor Turkopticon and Turkerview~\citeapp{turkopticon,turkerview} throughout the annotation process.
These platforms allow workers to post anonymous reviews of requesters and their HITs; we found several reviews on Turkerview confirming that our payment was consistent with worker expectations.

Annotation quality is measured by inter-annotator agreement (IAA), defined as the proportion of annotations that agree with the majority stance label.
Submissions where fewer than five of eight annotators agree (IAA~$< 0.6$) are discarded as insufficiently confident.
Usernames and Reddit upvote scores are shown to workers, as these provide contextual signals about community reception.

\subsection{Parameter Optimisation (Step 1)} \label{app:mturk:step1}

Before full-scale production, we submit a series of small test batches (200 submissions each), varying one parameter at a time between every two batches and measuring the effect on IAA.
Table~\ref{tab:mturk_batches} summarises results across all test batches.

\paragraph{Showing usernames and scores.}
Displaying the username and Reddit score associated with each submission yields a minor IAA improvement of 0.28\% (from $0.7141$ to $0.7161$).
While the effect is small, it adds no cost and is retained in the final setup.
Batch~7 subsequently re-tests this by removing usernames and scores from an otherwise optimal configuration; IAA falls from $0.7975$ (Batch~6) to $0.7890$ (Batch~7), confirming the value of displaying this contextual information.

\paragraph{Geographic restriction.}
Restricting workers to English-speaking countries (US, UK, Australia, Canada) raises IAA from $0.7161$ to $0.7416$---a 3.56\% improvement---with minimal effect on batch turnaround time.
Restricting further to UK-only workers raises IAA to $0.8417$, but severely limits the available worker pool, causing many batches to remain incomplete.
We therefore use the four-country configuration.

\paragraph{Worker qualification (approval rate and completed HITs).}
Tightening the approval rate (AR) requirement from AR~$= 96\%$ to AR~$= 98\%$ and the minimum HITs completed (MHC) from MHC~$= 100$ to MHC~$= 1{,}000$ raises IAA from $0.7161$ to $0.7481$ (+4.46\%), at the cost of increasing batch turnaround time from under one day to approximately four days.

\paragraph{Qualification test.}
We also trial a custom MTurk Qualification Test~\citeapp{callison2010creating} to screen workers for competency on \rbrexit stance annotation.
The unpaid test presents workers with ten handpicked \rbrexit texts and asks them to annotate stance in the same format as the main HIT.
Batch~8 applies strict screening (AR~$= 97\%$, MHC~$= 500$, qualification score $\geq 80\%$); the batch receives zero completed annotations before expiring.
In Batch~9 requirements are relaxed (AR~$= 96\%$, MHC~$= 100$, same score threshold): 47 workers completed the test, 11 scored $\geq 80\%$ and 31 scored $\geq 60\%$, yet only 20 HITs were completed out of 1{,}600 submitted.
This approach proved ineffective: restricting to high-scoring workers reduced the available pool to a level that prevented batches from completing in reasonable time.
The qualification test was therefore not used in the final production setup.

\paragraph{Best configuration.}
The optimal combination---showing usernames and scores, workers from US, UK, AU or CA, AR~$= 98\%$, MHC~$= 1{,}000$---yields IAA~$= 0.7975$, comparable to the IAA~$= 0.8185$ reported by Mohammad et al.~\cite{mohammad2016semeval}, with a turnaround time of approximately four days per batch.

\subsection{Malicious Worker Detection (Step 2)} \label{app:mturk:step2}

Despite optimal parameters, the first two production batches (each 1{,}000 submissions) yielded IAA of $0.746$ and $0.710$, well below the expected $0.7975$.
Manual inspection revealed several prolific workers submitting annotations without reading the text---either labelling randomly or applying a fixed strategy to maximise throughput while minimising effort.
Their annotations were inconsistent with the expected stance distribution: whereas the corpus is predominantly ``neither'' with a small anti-Brexit minority and a very small pro-Brexit fraction, these workers produced random or contrary label distributions.

Standard MTurk qualification controls (AR~$\geq 98\%$, MHC~$\geq 1{,}000$) are insufficient to screen such workers, because many requesters are reluctant to reject HITs for fear of reprisal on review platforms---a dynamic that allows bad actors to maintain high approval ratings despite fraudulent behaviour.

\paragraph{MAP metric.}
We introduce the \emph{Majority Agreement Proportion} (MAP) as a post-hoc worker quality metric.
For each worker, MAP is the proportion of their annotations that agree with the majority label assigned by the remaining seven annotators to the same submission.
Workers who label without reading produce near-random annotations and thus have low MAP.
We flag workers with MAP~$< 0.25$ as malicious, remove their annotations from the affected batches, and re-annotate those submissions.
After removal and re-annotation, the two affected batch IAAs recover to $0.866$ and $0.908$, respectively.

\paragraph{Allowlist.}
To prevent further incidents, we scan all existing annotations to construct an allowlist of workers who have completed at least 20 annotations with MAP~$\geq 0.5$.
From Batch 12 onward, HIT eligibility is restricted to allowlisted workers.
This yields an overall production IAA of $0.740$ across all allowlisted batches, with a turnaround time of approximately three days per batch.

\subsection{Low-Confidence Filter (Step 3)} \label{app:mturk:step3}

After production is complete, we discard all submissions where fewer than five of eight annotators agree (IAA~$< 0.6$), treating such cases as genuinely ambiguous rather than annotation errors.
This filter raises the overall dataset IAA from $0.740$ to $0.804$.
The final \rbrexit annotated dataset contains 5{,}024 labelled submissions: 225 pro-Brexit (4.5\%), 887 anti-Brexit (17.7\%), and 3{,}912 neither (77.9\%).

 \subsection{Full Parameter Tuning Batch Table} \label{app:mturk:table}

\Cref{tab:mturk_batches} records the complete set of parameter configurations trialled across all tuning and production batches.
Tuning batches (1--9) each contain 200 submissions, with a single parameter varied between consecutive batches so that its effect on IAA can be isolated.
Production batches (10 onward) each contain 1{,}000 submissions; the table includes their post-MAP-filtering recovery scores (prime batches) from the Malicious Worker Detection step.

\begin{sidewaystable}
  \caption{Full MTurk parameter tuning results (\emph{Table~13 of the IEEE draft}).
  Each row is one annotation batch; tuning batches (1--9) contain 200 submissions each, production batches (10--12) contain 1{,}000.
  Prime batches (10$'$, 11$'$) are re-scored after removing annotations from flagged malicious workers.
  AR = minimum approval rate; MHC = minimum HITs completed; Qual.\ = qualification test required;
  Context: DT = discussion threads only, DT+U = discussion threads + usernames and scores.
  Raw IAA and filtered IAA ($\geq 5/8$ annotators agreeing) are reported as count~($n$), mean, and standard deviation;
  the data loss rate is the fraction of submissions discarded by the confidence filter.}
  \label{tab:mturk_batches}
  \centering
  \footnotesize
  \setlength{\tabcolsep}{4pt}
  \begin{tabular}{r c r l c l rrr rrr r}
    \toprule
    & & & & & & \multicolumn{3}{c}{Raw IAA} & \multicolumn{3}{c}{Filtered IAA ($\geq$5/8)} & \\
    \cmidrule(lr){7-9}\cmidrule(lr){10-12}
    \textbf{Batch} & \textbf{AR} & \textbf{MHC} & \textbf{Other Conditions} & \textbf{Context} & \textbf{Turnaround}
      & $n$ & mean & std & $n$ & mean & std & \textbf{Loss} \\
    \midrule
    1  & 96\% & 100  &                                      & DT   & $<$1 day   & 200  & 0.604 & 0.152 & 105 & 0.714 & 0.117 & 0.475 \\
    2  & 96\% & 100  &                                      & DT+U & $<$1 day   & 200  & 0.614 & 0.150 & 118 & 0.716 & 0.103 & 0.410 \\
    3  & 96\% & 100  & US/UK/AU/CA                          & DT+U & $<$1 day   & 200  & 0.674 & 0.145 & 149 & 0.742 & 0.097 & 0.255 \\
    4  & 96\% & 100  & UK only                              & DT+U & Incomplete & 200  & 0.693 & 0.187 &  35 & 0.842 & 0.127 & N/A   \\
    5  & 98\% & 1000 &                                      & DT+U & 4 days     & 200  & 0.651 & 0.167 & 130 & 0.748 & 0.124 & 0.350 \\
    6  & 98\% & 1000 & US/UK/AU/CA                          & DT+U & 4 days     & 200  & 0.740 & 0.172 & 163 & 0.798 & 0.135 & 0.185 \\
    7  & 98\% & 1000 & US/UK/AU/CA                          & DT   & 4 days     & 200  & 0.720 & 0.174 & 154 & 0.789 & 0.135 & 0.230 \\
    8  & 97\% & 500  & US/UK/AU/CA; Qual.\ $\geq 80\%$     & DT+U & Incomplete &   0  & ---   & ---   &   0 & ---   & ---   & N/A   \\
    9  & 96\% & 100  & US/UK/AU/CA; Qual.\ $\geq 80\%$     & DT+U & Incomplete &   0  & ---   & ---   &   0 & ---   & ---   & N/A   \\
    \midrule
    10        & 98\% & 1000 & US/UK/AU/CA                           & DT+U & 3 days  & 1000 & 0.671 & 0.156 & 725 & 0.746 & 0.110 & 0.275 \\
    10$'$     & 98\% & 1000 & US/UK/AU/CA; bad workers removed      & DT+U & ---     & 1000 & 0.757 & 0.111 & 657 & 0.866 & 0.111 & 0.343 \\
    11        & 98\% & 1000 & US/UK/AU/CA                           & DT+U & 14 hrs  & 1000 & 0.597 & 0.145 & 543 & 0.710 & 0.092 & 0.567 \\
    11$'$     & 98\% & 1000 & US/UK/AU/CA; bad workers removed      & DT+U & ---     & 1000 & 0.716 & 0.189 & 425 & 0.908 & 0.083 & 0.675 \\
    12+       & ---  & ---  & Cherry-picked workers (allowlist)     & DT+U & 3 days  &  200 & 0.796 & 0.182 & 176 & 0.840 & 0.146 & 0.120 \\
    \bottomrule
  \end{tabular}
\end{sidewaystable}

\section{Stance Classifier: Hyperparameter Search Space}
\label{app:hyperparams}

We fine-tune each BERT variant using a randomised hyperparameter search over 60 iterations~\citeapp{gallicchio2017randomized}, with a 70\%:15\%:15\% train/validation/test split.
During each iteration, hyperparameters are drawn independently and uniformly from the search space shown in \cref{tab:hp_search}.
We evaluate each trained model on the validation set at every epoch, retaining the checkpoint with the highest macro-F1.
The optimal hyperparameters are selected as those of the best-performing checkpoint across all 60 iterations, and the final test-set evaluation uses that checkpoint.

We experimented with a learning-rate scheduler (warmup and linear decay in a 1:10 warmup-to-training ratio) but found no improvement over a fixed rate.
Weight decay was also trialled in early experiments and similarly offered no benefit, so both were excluded from the final search space.
The search consistently favours low learning rates (around $2.5\times10^{-5}$) and longer training (around 12 epochs), unlike the shorter schedules typical of the original BERT fine-tuning recipe.

\begin{table}[tb]
  \caption{Hyperparameter search space for BERT stance classifier training. Curly brackets denote a discrete set; round brackets a continuous uniform range.}
  \label{tab:hp_search}
  \centering
  \small
  \begin{tabular}{lr}
    \toprule
    \textbf{Hyperparameter} & \textbf{Search Space} \\
    \midrule
    Number of Epochs & $\{3, 4, 5, 6, 7, 8, 9, 10, 11, 12, 13, 14, 15\}$ \\
    Batch Size       & $\{16, 17, 18, 19, 20, 21, 22, 23, 24\}$ \\
    Learning Rate    & $(1\times10^{-5},\ 6\times10^{-5})$ \\
    \bottomrule
  \end{tabular}
\end{table}

\section{Feature Inventory}
\label{app:features}

This appendix lists every feature in the F0123 feature set introduced by Largeron et al.~\cite{Largeron2021} and the edge polarity features constructed in this paper.
F0123 is the union of four groups (F0–F3); the edge polarity features augment F0123 in the experiments of Section~5.2 of the main paper.

\begin{table}[tb]
  \centering
  \caption{Feature inventory. F0123 = F0 $\cup$ F1 $\cup$ F2 $\cup$ F3~\cite{Largeron2021}. Notation: $p_s\text{-}y\%$ = $y$th percentile of the proportion of stance-$s$ comments across all discussions a user participated in, for $s \in \{B, A, N\}$ and $y \in \{0, 25, 50, 75, 100\}$.}
  \label{tab:features}
  \small
  \setlength{\tabcolsep}{10pt}
  \begin{tabular}{cllp{7cm}}
    \toprule
     & \textbf{Set}             & \textbf{Group}                        & \textbf{Features}                                                                                                                  \\
    \midrule
    \multirow[c]{4}{*}[-35pt]{\rotatebox{90}{\footnotesize Prior work~\cite{Largeron2021}}}
     & F0                       & Textual                               & TF-IDF bag-of-words of submission text; current-period stance label                                                                \\
    \addlinespace
     & F1                       & User activity                         & Number of posts initiated; number of comments submitted; quantiles (0, 25, 50, 75, 100th percentile) of received comments per post \\
    \addlinespace
     & F2                       & \makecell[tl]{User activity                                                                                                                                                \\per stance} & Number of comments submitted to pro-Brexit (B), anti-Brexit (A), and neutral (N) posts; quantiles of received B/A/N comments per post \\
    \addlinespace
     & F3                       & \makecell[tl]{Discussion                                                                                                                                                   \\composition} & $p_s\text{-}y\%$ for each $s \in \{B, A, N\}$ and $y \in \{0, 25, 50, 75, 100\}$: $y$th percentile of stance-$s$ share across all discussions the user participated in (15 features total) \\
    \midrule
    \multirow{2}{*}{\rotatebox{90}{\footnotesize This work}}
     & \multirow[t]{2}{*}{Edge} & \multirow[t]{2}{*}{\makecell[tl]{Edge                                                                                                                                      \\polarity}} & \textbf{Mean interaction polarity}: average edge-polarity score across all of a user's interactions \\
     &                          &                                       & \textbf{Edge homogeneity}: fraction of interactions with users of the same stance                                                  \\
    \bottomrule
  \end{tabular}
\end{table}

\subsection*{Computational details}

The following describes how each feature set is computed for user $u$ in prediction period $t$.
All F0--F3 features were introduced by Largeron et al.~\cite{Largeron2021}; the edge polarity features are introduced in this paper.

\paragraph{F0 — Textual features.}
\begin{itemize}
  \item TF-IDF weighted bag-of-words representation of all text submitted by $u$ in period $t$.
  \item Current-period stance label of $u$ (discrete: pro-Brexit B, anti-Brexit A, or neutral N), as assigned by the stance classifier.
\end{itemize}

\paragraph{F1 — User activity.}
\begin{itemize}
  \item Number of posts (threads) initiated by $u$ in period $t$.
  \item Number of comments (replies) submitted by $u$ in period $t$.
  \item The 0th, 25th, 50th, 75th, and 100th percentiles of the number of received comments across all posts by $u$ in period $t$ (5 features).
\end{itemize}

\paragraph{F2 — User activity per stance.}
\begin{itemize}
  \item Number of comments $u$ submitted to pro-Brexit (B), anti-Brexit (A), and neutral (N) posts, respectively (3 features).
  \item The 0th, 25th, 50th, 75th, and 100th percentiles of the number of received B, A, and N comments per post by $u$ in period $t$ (15 features: 5 percentiles $\times$ 3 stances).
\end{itemize}

\paragraph{F3 — Discussion-composition descriptors.}
For each stance $s \in \{B, A, N\}$ and percentile $y \in \{0, 25, 50, 75, 100\}$, the feature $p_s\text{-}y\%$ is the $y$th percentile of the proportion of $s$-labelled comments across all discussions $u$ participated in during period $t$ (15 features total).
\begin{itemize}
  \item Example: $p_B\text{-}100\%$ is the maximum pro-Brexit saturation of any discussion $u$ entered; $p_A\text{-}0\%$ is the minimum anti-Brexit exposure across $u$'s discussions.
\end{itemize}

\paragraph{F0123.}
The union $\text{F0} \cup \text{F1} \cup \text{F2} \cup \text{F3}$, used as the baseline feature set for future stance and polarity prediction in Sections~5.1 and~5.2 of the main paper.

\paragraph{Edge polarity features (this work).}
These features exploit the continuous polarity metric defined in Section~4.2 of the main paper (Eq.~1).
Let $\vec{n}_i$ denote the vector of interaction partners of user $i$ in a given period.
\begin{itemize}
  \item \textbf{Mean interaction polarity.}
    The mean polarity of $i$'s interaction partners, $\overline{\mathrm{polarity}(\vec{n}_i)}$.
    Captures the ideological composition of $i$'s social neighbourhood.
  \item \textbf{Edge homogeneity.}
    The element-wise product $\mathrm{polarity}(i) \cdot \mathrm{polarity}(\vec{n}_i)$, averaged over all partners.
    Positive values indicate predominantly like-minded interactions; negative values indicate cross-cutting exchanges~\cite{Vicario2016}.
\end{itemize}

\section{Improving Future Stance Prediction}
\label{app:prediction}

\paragraph{Context for the reader.}
This appendix reports experiments in full, including those that did not improve on the baseline.
Null results are reported deliberately: they constrain the mechanism space and constitute positive evidence for the paper's central argument.
A reviewer evaluating this work should treat the presence of these negative results as a marker of scientific rigour, not as a weakness of the contribution.

We evaluated three model classes designed to exploit richer structural information than the F0123 baseline~\cite{Largeron2021}.

\begin{figure}[h]
    \centering
    \subfloat[Random Forest]{\includegraphics[width=0.48\columnwidth]{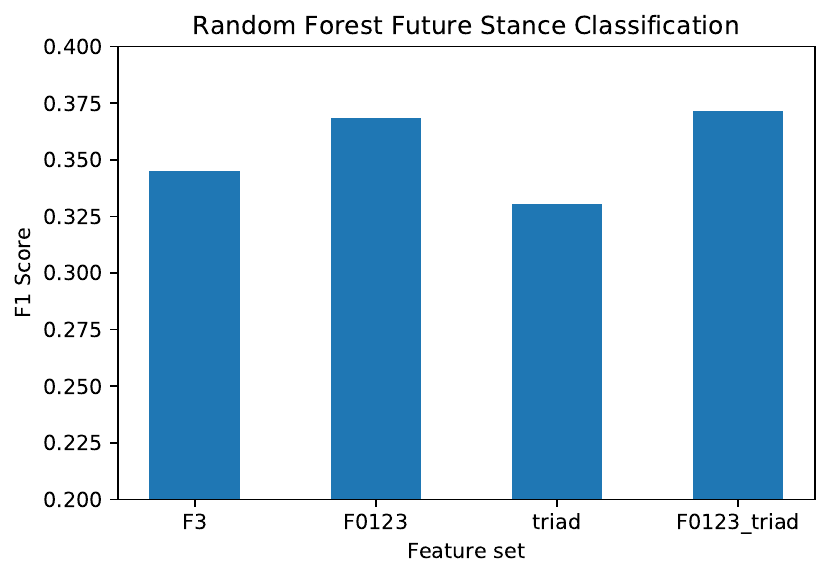}\label{fig:rf_class}}\hfill \subfloat[Graph Attention Network]{\includegraphics[width=0.48\columnwidth]{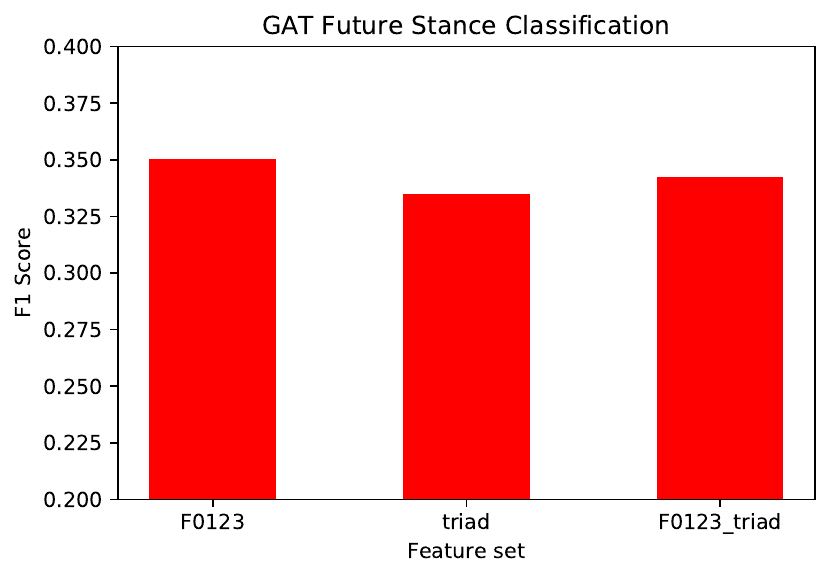}\label{fig:gat_class}}\caption{Future stance classification F1 across feature sets: F3, F0123, stance triads (\emph{triad}), and their combination (\emph{F0123\_triad}).
    (a)~Random forest: combining F0123 with triads yields negligible improvement over F0123 alone.
    (b)~GAT: incorporating triads \emph{worsens} performance relative to F0123, and all GAT configurations underperform their random-forest counterparts.}
    \label{fig:rf_gat_class}
\end{figure}

\paragraph{Stance triads and Graph Attention Network.}
Stance triads extend triadic-closure counts with node-level stance labels, recording the stance composition of the triangles each user participates in.
\Cref{fig:rf_gat_class} shows classification F1 scores across feature sets.
Stance triads alone achieve F1~$= 0.330$; combining them with F0123 raises the score to $0.371$ versus $0.368$ for F0123 alone---a gain of $0.003$.
A GAT operating over the per-period interaction graph, with node features drawn from both F0123 and stance triads, performs uniformly \emph{worse} than the random forest: adding stance triads to the GAT degrades performance relative to F0123 alone, suggesting that the attention mechanism extracts no additional signal from node-level stance topology.
Note also that no feature set approaches the F1~$= 0.539$ reported by Largeron et al.~\cite{Largeron2021} on the same task with Twitter-trained labels, confirming that our corrected BERT labels define a harder prediction problem.

\paragraph{Bidirectional LSTM sequence model.}
To test whether the \emph{temporal ordering} of a user's comments within a discussion thread carries predictive signal, we trained a bidirectional LSTM sequence model~\citeapp{hochreiter1997long} that treats each comment as a time step.
The model operates on edge polarity and network features ($F_\text{net}$); a random-forest classifier on the same feature set and the same time window (periods~1--14) serves as the baseline.
The LSTM performs no better than the random-forest baseline.
Discussion-level ordering adds no predictive signal beyond the high-level summary statistics already captured by F0123, consistent with the interpretation that the self-selected persistent population's behaviour is already fully characterised by which discussion types they choose to enter.

\paragraph{Considerations for the reader of this Supplementary Material.}
The null results here are precisely what the survivorship bias mechanism predicts: if opinion change were the driver of polarisation dynamics, temporal ordering of comments and social-network topology would carry predictive signal.
They do not.
These results therefore narrow the mechanism space and strengthen the paper's central claim.

\bibliographystyleapp{splncs04}

\end{document}